\documentclass[12pt]{article}
\usepackage{amsmath}
\usepackage{graphicx}
\usepackage{enumerate}
\usepackage{natbib}
\usepackage[font=tiny,labelfont=bf]{caption}
\usepackage{url} % not crucial - just used below for the URL 
\usepackage{bm}
\usepackage{tikz}
\usepackage{tikz-network}
\usepackage{pgfplots}
\usepackage{graphicx}
\usepackage{amsmath}
\usepackage{amsbsy}
\usepackage{amssymb}
\usepackage{amsfonts}
\usepackage{mathtools}
\usepackage{booktabs}
\usepackage{floatrow}
\usepackage{subfig}
 \usepackage{relsize}
\usepackage{array}
\usepackage{comment}
\usepackage{algorithm,algcompatible,amsmath}
\algnewcommand\INPUT{\item[\textbf{Input:}]}%
\algnewcommand\OUTPUT{\item[\textbf{Output:}]}%
\usepackage{xr}
\usepackage{cleveref}
\usepackage{multirow}
\usepackage{amsthm}
\usepackage{makecell}

\usepgfplotslibrary{groupplots,dateplot}
\usetikzlibrary{patterns,shapes.arrows}
\usetikzlibrary{arrows.meta}
\pgfplotsset{compat=1.17}
\newcommand{\Fnorm}[1]{\left\lVert#1\right\rVert_F}

\DeclareUnicodeCharacter{2212}{−}
\definecolor{cerulean}{rgb}{0.0, 0.48, 0.65}
\newtheorem{thm}{Theorem}%[section]

 \usepackage{xr}
\usepackage{cleveref}
\usepackage{algorithm}
\usepackage{algpseudocode}

\def\tA{\mathcal{A}}
\def\tB{\mathcal{B}}

\def\tE{\mathcal{E}}
\def\tF{\mathcal{F}}

\def\tS{\mathcal{S}}
\def\tT{\mathcal{T}}

\def\tY{\mathcal{Y}}

\newcommand{\twonorm}[1]{\left\lVert#1\right\rVert_2}

\newcommand{\zeronorm}[1]{\left\lVert#1\right\rVert_0}

\newtheorem{assumption}{Assumption}

\newtheorem{condition}{Condition}

\crefname{figure}{}{}
\crefname{table}{}{}
\crefname{equation}{}{}
\crefname{thm}{}{}
\crefname{lem}{}{}
\crefname{algorithm}{}{}
\crefname{section}{}{}

\newcommand{\blind}{1}
\begin{document}

\def\spacingset#1{\renewcommand{\baselinestretch}%
{#1}\small\normalsize} \spacingset{1}

%%%%%%%%%%%%%%%%%%%%%%%%%%%%%%%%%%%%%%%%%%%%%%%%%%%%%%%%%%%%%%%%%%%%%%%%%%%%%%

\if1\blind
{
  \title{\bf Sparse higher order partial least squares for
simultaneous variable selection, dimension reduction, and tensor denoising}

\author{Kwangmoon Park$^1$\\ kpark243@wisc.edu \and S\"und\"uz Kele\c{s}$^1$$^2$ \thanks{Corresponding author}\\ keles@stat.wisc.edu}
\date{%
    $^1$Department of Statistics, University of Wisconsin, Madison, WI, USA, 53706\\%
    $^2$Department of Biostatistics and Medical Informatics, University of Wisconsin, Madison, WI, USA, 53726\\[2ex]%
    \today
}

\maketitle

} \fi

\if0\blind
{
  \bigskip
  \bigskip
  \bigskip
  \begin{center}
    {\LARGE\bf Sparse higher order partial least squares for simultaneous variable selection, dimension reduction, and tensor denoising}
\end{center}
  \medskip
} \fi

\bigskip
\begin{abstract}
Partial Least Squares (PLS) regression emerged as an alternative to ordinary least squares for addressing multicollinearity in a wide range of scientific applications. As multidimensional tensor data is becoming more widespread, tensor adaptations of PLS have been developed. In this paper, we first establish the statistical behavior of Higher Order PLS (HOPLS) of \cite{zhao2012higher}, by showing that the consistency of the HOPLS estimator cannot be guaranteed as the tensor dimensions and the number of features increase faster than the sample size. To tackle this issue, we propose \textbf{S}parse \textbf{H}igher \textbf{O}rder \textbf{P}artial Least \textbf{S}quares (SHOPS) regression and an accompanying algorithm. SHOPS simultaneously accommodates variable selection, dimension reduction, and tensor response denoising. We further establish the asymptotic results of the SHOPS algorithm under a high-dimensional regime. The results also complete the unknown theoretic properties of SPLS algorithm \citep{chun2010sparse}. We verify these findings through comprehensive simulation experiments, and application to an emerging high-dimensional biological data analysis. 

\end{abstract}

\noindent%
\vfill

\newpage
\spacingset{1.9} % DON'T change the spacing!

\section{Introduction}

Recent developments in biotechnology such as single cell sequencing and brain imaging enabled collecting  high-dimensional and complex structured biological data. For example, data from sequencing experiments often  yields numbers of  features ($p$) that readily outnumber the sample sizes ($n$), i.e., $n\ll p $. 
In addition, multi-dimensional arrays, also known as higher order tensors, naturally arise in these complex experiments from the fields of genomics, neuroscience, and chemometrics, and harbor 
more complicated characteristics compared to classical lower order/dimensional data. 

In the high-dimensional regression settings, the two key issues, namely selection of important variables and reduction of multicolllinearity,  have been extensively studied over the last three decades. The former is largely addressed by leveraging the principles of sparsity where a subset of the variables derive the signal \citep{tibshirani1996regression,fan2001variable,zou2003regression,yuan2006model,raskutti2011minimax,hastie2015statistical,li2017feature}. For the latter case, dimension reduction-based approaches including Principal Component Analysis (PCA)  and Partial Least Squares (PLS)-based methods \citep{wold1966estimation,de1993simpls,helland1994comparison} are widely adopted. Among these dimension reduction approaches, Sparse Partial Least Squares (SPLS) model \citep{chun2010sparse}  addresses both of the challenges in highly correlated, high-dimensional data settings and performs superior to  non-sparse PLS which has lower estimation accuracy and less practical interpretability  in the high-dimensional regimes.

With the availability of multi-dimensional array (tensor) structured data in many scientific fields, a number of tensor focused methods to address tensor regression problems has emerged \citep{zhao2012higher,zhang2017tensor,li2017parsimonious,sun2017store,lock2018supervised,raskutti2019convex}. Among these, tensor PLS or envelop regression models \citep{zhao2012higher,zhang2017tensor,li2017parsimonious} focused on tensor coefficient denoising with reduction of high-dimensional coefficients, and a subset of them established  asymptotic results for the PLS models under large-sample scenario (\textbf{Tab.}~\cref{supp-tab:summaryTensor}). However, the resulting PLS direction vectors of these methods are linear combinations of all the variables, potentially obscuring the interpretation. Moreover, when the response tensor does not harbor any real association with the covariates, non-sparse tensor PLS methods cannot avoid estimating redundant associations. While these issues seem to hinder both interpretation and estimation accuracy, the statistical behavior of tensor PLS is still unknown under the high-dimensional regime, where both the number of features and the response tensor dimension grow faster than the sample size. 

The main contributions of this paper are three-folds. \textbf{1)} We establish that the tensor PLS coefficient estimator can be not only hard to interpret, but also inconsistent in the high-dimensional regime, which emerges in numerous scientific domains including biology, chemometrics, and neuroscience. In providing theoretic guarantees for HOPLS, we carefully explain how the PLS assumptions for matrices, which are employed in \cite{chun2010sparse}, naturally extend to the assumptions we impose for the extension to higher order cases. \textbf{2)} To address these key issues, we develop a Sparse Higher Order Partial least Squares (SHOPS) model  that aims to (i) conduct simultaneous variable selection and denoising in-active entries of the response tensor and  (ii) perform dimension reduction that best summarizes the association between the covariates and the response (\textbf{Fig.}~\ref{fig:intro}a). To the best of our knowledge, in the PLS domain, this paper is the first  to suggest tensor response partial least squares method with explicit sparsity assumption on the coefficient tensor (\textbf{Tab.}~\cref{supp-tab:summaryPLS}). \textbf{3)} We demonstrate statistical properties of SHOPS under the high-dimensional regime, which also address the unresolved theoretical guarantee of the both univariate and multivariate SPLS algorithm \citep{chun2010sparse}, one of the most recognized methods among the PLS-type methods.

In what follows, we first review tensor response partial least squares model and establish an unknown asymptotic result in the high-dimensional regime.
%that shows that classical low-dimensional consistency result does not hold when the number of parameters exceeds the sample size. In Section 3, we propose the SHOPS regression model and an accompanying algorithm for estimation, and establish its asymptotic properties. Specifically, we show that SHOPS estimator achieves consistency even under the high-dimensional paradigm. In Section 4, we conduct a simulation study and compare SHOPS with a variety of baseline multivariate (or tensor) response methods. Section 5 showcases the utility of the SHOPS model with a high-dimensional biological application to the mouse brain single cell high-throughput chromatin conformation capture (scHi-C) data integrated with single cell DNA methylation  data.

\section{Tensor partial least squares}
\label{sec:TPLS}

\subsection{Notation and preliminaries}
We begin with a brief review of the notations of the  tensor framework  \citep{kolda_review}. The $(d_1,d_2,\cdots, d_M)$-dimensional, real valued array $\mathcal{Y}\in \mathbb{R}^{d_1\times d_2\times \cdots\times d_M}$ is a $M$th order tensor, which has $M$ \textit{modes} of  dimension. For a third order tensor $\mathcal{Y}\in \mathbb{R}^{d_1\times d_2\times d_3}$ and a matrix $A\in \mathbb{R}^{r\times d_1}$, we define the tensor multiplication 
along the first mode of the tensor, $\times_1 : \mathbb{R}^{d_1\times d_2\times d_3} \times \mathbb{R}^{r\times d_1}\rightarrow  \mathbb{R}^{r\times d_2\times d_3}$, as $\mathcal{Y}\times_1 A=\left(\sum_{i'=1}^{d_1}{y_{i'jk}a_{ii'}}\right)_{1\leq i\leq r,
1\leq j\leq d_2,1\leq k\leq d_3}.$ Mode $m$ multiplication operators $\times_m$ are also defined similarly. The Frobenius norm of a tensor $\mathcal{Y}\in \mathbb{R}^{d_1\times d_2\times\cdots\times d_M}$ is defined as the square root of the sum of squares of the entire set of entries. For example, for the third order tensor, i.e., $M=3$, we have $\Fnorm{\mathcal{Y}}=\sqrt{\sum_{i=1}^{d_1}\sum_{j=1}^{d_2}\sum_{k=1}^{d_3}y_{ijk}^2}.$
In addition, we define the spectral norm of a tensor $\tT\in \mathbb{R}^{d_1\times \cdots \times d_M}$ as $\twonorm{\tT}=\max\limits_{\bm{u}_{m}\in \mathbb{S}^{d_m-1}}\{\tT\times_1 \bm{u}_{1}^T\cdots\times_M \bm{u}_{M}^T\}$. For a matrix $U\in \mathbb{R}^{p\times R}$, we also define $\zeronorm{U}\coloneqq \sum_{i=1}^p{1_{\{U_{[i,]}\neq 0\}}}$, which quantifies the number of non-zero rows. Operations $\otimes$ and  $\circ$ denote the Kronecker and the outer product, respectively. 

We define the unfolded matrix of a third order tensor $\mathcal{Y}\in \mathbb{R}^{d_1\times d_2\times d_3}$ along the $k$th mode as $\tY_{(k)}\in\mathcal{R}^{(\prod_{i\neq k}d_i)\times d_k}$. Then, the $(i,j,k)$th element of a tensor $\mathcal{Y}\in \mathbb{R}^{d_1\times d_2\times d_3}$ can be represented as an element of the relevant unfolded matrix along  each mode as $y_{ijk}=\{\tY_{(1)}\}_{(j-1)d_3+k,i}=\{\tY_{(2)}\}_{(k-1)d_1+i,j}
=\{\tY_{(3)}\}_{(i-1)d_2+j,k}.$ In this paper's mathematical representation, a lowercase letter in boldface, e.g., $\bm{v}$, denotes a vector, a capital letter, in general, is used for a matrix, and a  calligraphic letter, e.g., $\mathcal{Y},\mathcal{M}$,  is used for a tensor, unless otherwise specified. Note that $\mbox{superdiag}\left(a_j
\right)_{j=1}^K\in\mathbb{R}^{K\times \cdots\times K}$ refers to a tensor whose each $(j,j,\cdots,j)$ entry is filled with $a_j$. For a set of sequential numbers, i.e., $\{1,\cdots,K\}$, we use the abbreviation $[K]$ for simplicity. We adopt R programming language convention to denote tensor extraction, unless specified otherwise. For example, for sets $\tA_2\subset[d_2]$, $\tA_3\subset[d_3]$, $\tT_{[\, \, ,\tA_2,\tA_3]}\in \mathbb{R}^{d_1\times s_2\times s_3}$ denotes the extraction of $\tA_2$ and $\tA_3$ entries of the second and the third mode of the original tensor $\tT\in \mathbb{R}^{d_1\times d_2\times d_3}$, where $s_1,s_2$ denotes the cardinality of the sets $\tA_1$ and  $\tA_2$, respectively. 

Next section formulates the ordinary tensor Partial Least Squares model by utilizing the notations defined in this section.

\subsection{Tensor partial least squares model}\label{subsec:TPLSmodel}

The tensor Partial Least Squares (PLS) model, also known as the Higher Order Partial Least Squares (HOPLS), was introduced by \cite{zhao2012higher}. 
%Multiple variants of the tensor PLS models have emerged over the years \citep{zhao2012higher, zhang2017tensor, li2017parsimonious}. 
Here, motivated by %the two 
a high-dimensional genomic application, we consider the case where the response is a higher order tensor $\tY\in \mathbb{R}^{n\times d_2\times d_3\cdots,\times d_M}$ and the covariates form a matrix, i.e., $X\in\mathbb{R}^{n\times p}$ with $p=d_1$. Withoug loss of generality, we focus on the case $M=3$ although the extension to higher $M$ is straightforward. Note that this PLS is still a higher-order PLS ($M\geq 3$), as the coefficient tensor is with order three, and the entire theoretic analyses and algorithm implementation require arguments involving higher order tensor objects. In addition, we mainly focus on the NIPALS formulation \citep{wold1966estimation} of the PLS model, as the SPLS-NIPALS performs generally better than its SIMPLS formulation \citep{de1993simpls} of SPLS in \cite{chun2010sparse}.

 When the regression problem is ill-posed, i.e., $n<p$,  the key task of the PLS is to find a solution for the regression coefficients that incorporates the relation between $X$ and $\tY$ the most by finding subspace of the covariance tensor $\Sigma_{X\tY}$. This is comparable to another popular choice of dimension reduction-based regression method, Principal Component Regression, in that it finds the subspace of  the covariance matrix $\Sigma_{X}$.

The NIPALS \citep{wold1966estimation} version of tensor response PLS model for $\tY\in \mathbb{R}^{n\times d_2\times d_3}$ and $X\in \mathbb{R}^{n\times d_1}$ can be formulated as
\begin{align*}
\tY&=\tB\times_1X +\sigma_2\tF\\
&=\tB_W\times_1XW +\sigma_2\tF,
\end{align*}
where $\sigma_2>0$ is a scalar and $\tF$ denotes the error tensor. Note that 
\begin{equation*}
\tB_W=\mathcal{G}\times_1(W^T\Sigma_{X}W)^{-1}\times_2 Q_{2}\times_3 Q_{3}    
\end{equation*}
with $W(=W_K)=[\bm{w}_1\cdots,\bm{w}_K]\in \mathbb{R}^{p\times K}$, $Q_2=[\bm{q}^{(2)}_1\cdots,\bm{q}^{(2)}_K]\in \mathbb{R}^{d_2\times K}$, $Q_3=[\bm{q}^{(3)}_1\cdots,\bm{q}^{(3)}_K]\in \mathbb{R}^{d_3\times K}$, and 
$\mathcal{G}=\mbox{superdiag}\left(g_k
\right)_{k=1}^K\in\mathbb{R}^{K\times K\times K}$, where $g_k= \Sigma_{X\tY}
\times_1 \bm{w}_k^T\times_2 \bm{q}_k^{(2)T}\times_3 \bm{q}_k^{(3)T}$. Furthermore,
for population covariances $\Sigma_{X\mathcal{Y}}=\mathbb{E}[\mathcal{Y}\times_1 X^T/n]$, $\Sigma_{X}=\mathbb{E}[X^TX/n]$, the latent direction vectors $\bm{w}_{k},\bm{q}_k^{(2)},\bm{q}_k^{(3)}$ are the successive loading solutions of 
\begin{equation*}
\max\limits_{\bm{s}_1,\bm{s}_2,\bm{s}_3}{ |\Sigma_{X\mathcal{Y}}\times_1 \bm{s}
_1\times_2 \bm{s}_2\times_3 \bm{s}_3|^2 },    
\end{equation*}
subject to $\bm{s}_1^T\Sigma_{X}\bm{w}_j=0$  for $ j\in [k-1]$, and $\bm{s}_1^T(I_{d_1}-W_{k-1}W^+_{k-1})\bm{s}_1=1$, $\bm{s}_2\in\mathbb{S}^{d_2-1}$, $\bm{s}_3\in\mathbb{S}^{d_3-1}$. Here, $W_{k-1}^+$ denotes the Moore-Penrose inverse of $W_{k-1}=(\bm{w}_1,\cdots,\bm{w}_{k-1})$, and $\mathbb{S}^{p-1}$ denotes a unit sphere on $\mathbb{R}^p$. 
In this formulation, the latent direction vector $W$ plays a key role in conducting dimension reduction of high-dimensional $X$ in PLS by decomposing the covariance tensor $\Sigma_{X\tY}$. In addition, the latent directions along the other modes $Q_2,Q_3$ perform tensor response denoising. Note that $XW\in \mathbb{R}^{n\times K}$ corresponds to the PLS component matrix in a reduced dimension, and the tensor $\tB_W$ is the coefficient tensor for the PLS component. Once we have the estimates $\hat{W},\hat{Q}_2,\hat{Q}_3$, we can form the PLS coefficient tensor as
$\hat{\tB}_{PLS}=\hat{\mathcal{G}}\times_1\hat{W}(\hat{W}^TS_{X}\hat{W})^{-1}\times_2 \hat{Q}_{2}\times_3 \hat{Q}_{3}$, where $\hat{\mathcal{G}}=\mbox{superdiag}\left(\hat{g}_k
\right)_{k=1}^K$, $\hat{g}_k= S_{X\tY}
\times_1 \hat{\bm{w}}_k^T\times_2 \hat{\bm{q}}_k^{(2)T}\times_3 \hat{\bm{q}}_k^{(3)T}$ 
%$\hat{\mathcal{B}}_{PLS}=\mathcal{Y}\times_1\hat{W}(\hat{W}^TX^TX\hat{W})^{-1}\hat{W}^TX^T=S_{X\tY}\times_1\hat{W}(\hat{W}^TS_{X}\hat{W})^{-1}\hat{W}^T$, where 
and the sample covariance tensor is defined as $S_{X\tY}=\mathcal{Y}\times_1 X^T/n$ and the sample covariance matrix of $X$ is defined as $S_{X}=X^TX/n$.

The HOPLS can be implemented with Algorithm~\ref{alg:HOPLS}. Step 2 of the algorithm, where the rank-(1,1,1) decomposition on the covariance tensor is conducted, reveals the main difference against matricized response PLS in terms of finding the PLS direction vectors. Specifically, unless a low dimensional structure of the covariance tensor ($S_{X\tY_k}$)  is completely absent, the tensor PLS gives more accurate PLS estimate of the latent direction vectors $W$ than matrix PLS. 

\begin{algorithm}\footnotesize
\caption{Higher Order Partial Least Squares}\label{alg:HOPLS}
\begin{algorithmic}

\State \textbf{Initialize} ${\tY}_1=\tY$, ${X}_1=X$.
\For {$k=1,\cdots, K$}.
\State 1. Calculate $S_{X\tY_k}={\tY}_k\times_1{X}_k/n$.
\State 2. Conduct Rank$-(1,1,1)$ Higher Order Orthogonal Iteration \citep{de2000best} (HOOI) decomposition. i.e., obtain 1st, 2nd, 3rd loadings 
$ \bm{q}_k^{(1)},\bm{q}_k^{(2)},\bm{q}_k^{(3)}\leftarrow \mbox{HOOI}(S_{X\tY_k})$.

\State 3. Update $T_k=({X}_1\bm{q}_1^{(1)},\cdots, {X}_k\bm{q}_k^{(1)})$, $Q_{mk}=(\bm{q}_1^{(m)},\cdots, \bm{q}_k^{(m)})$ for $m=1,2,3$
\State 4. Update ${\tY}_{k+1}=\tY\times_1(I-T_kT_k^+)%\times_2 \bm{q}_k^{(2)}\bm{q}_k^{(2)T}\times_3 \bm{q}_k^{(3)}\bm{q}_k^{(3)T}
$, ${X}_{k+1}=(I-T_kT_k^+)X$.

\EndFor

\State Set $W_K=Q_{1K}(P_K^TQ_{1K})^{-1}$, where $P_K=X^TT_K(T_K^TT_K)^{-1}$, and \\$\mathcal{G}_K=\mbox{superdiag}\left(g_k
\right)_{k=1}^K\in\mathbb{R}^{K\times K\times K}$, where $g_k\coloneqq S_{X\tY}
\times_1 \bm{w}_k^T\times_2 \bm{q}_k^{(2)T}\times_3 \bm{q}_k^{(3)T}$\\
\textbf{Return} $\hat{\tB}=\mathcal{G}_K\times_1W_K(W_K^TS_{X}W_K)^{-1}\times_2 Q_{2K}\times_3 Q_{3K}$\\
(or equivalently, $\hat{\tB}=\sum_{k=1}^KS_{X\tY}\times_1\bm{w}_k(\bm{w}_k^{T}S_{X}\bm{w}_k)^{-1}\bm{w}_k^{T}\times_2 \bm{q}_k^{(2)}\bm{q}_k^{(2)T}\times_3 \bm{q}_k^{(3)}\bm{q}_k^{(3)T}$)
\end{algorithmic}
\end{algorithm}

\subsection{Asymptotic results for tensor partial least squares}

%The consistency of tensor partial least squares coefficient estimator is suggested in \cite{zhang2017tensor} in relation to the envelope tensor regression  for the case with  tensor covariate $\tX$ and  matrix/vector response; however, this result does not immediately extend to the tensor response setting we are considering here.  Furthermore, the asymptotic result is characterized only for $n\rightarrow \infty$, and not for the case when the dimensions of the coefficient tensor grows. 
%On the other hand, asymptotic results for tensor response envelope regression are investigated in \cite{li2017parsimonious} including asymptotic normality of the coefficient estimator. 
%However, in this paper, envelope of the response space is considered, instead of that of covariates $X$. Hence, the result cannot be discussed in a direct relation to the tensor response PLS estimation results. In addition, the asymptotic results are also for the classical regime when the sample size $n$ grows faster than the coefficient tensor dimension.  
Here, we first derive asymptotic properties of ordinary tensor partial least squares estimator when both the sample size $n$, and also the coefficient tensor size $d_1,d_2,d_3$ increase  at certain rates. We then argue that the classical consistency of the coefficient tensor estimator $\hat{\tB}$ can not be guarnateed in certain cases.

In order to derive asymptotic results for the tensor partial least squares, we assume that the model satisfies the following assumptions. Assumption \ref{asm1}, referred to as \textit{spiked covariance model}, is also employed in both \cite{johnstone2009consistency} and  \cite{chun2010sparse} for the covariance structure of the covariate matrix $X$. 

\begin{assumption} \label{asm1}
Assume that each row of $X=(\bm{x}_1^T,\cdots,\bm{x}_n^T)$ follows the model $\bm{x}_i=\sum_{j=1}^l v_j^i\bm{\gamma}_j+\sigma_1\bm{e}_i$, for some constant $\sigma_1$, where \textbf{(a)} $\bm{\gamma}_j$, $j=1\cdots l\leq p$, are mutually orthogonal PCs with norms $\twonorm{\bm{\gamma}_1}\geq\twonorm{\bm{\gamma}_2}\geq \cdots \geq \twonorm{\bm{\gamma}_l},$ \textbf{(b)} the multipliers $v_j^i\sim N(0,1)$ are independent over the indices of both $i$ and $j$, \textbf{(c)} the noise vectors $\bm{e}_i\sim N(0,I_p)$ are independent among themselves and of the random effects $\{v_i^j\}$ and \textbf{(d)} $p(n)$, $l(n)$, and $\{\bm{\gamma}_j(n),j=1,\cdots,m\}$ are functions of $n$, and the norms of PCs converge as sequences: $\varrho(n)=(\twonorm{\bm{\gamma}_1(n)},\cdots,\twonorm{\bm{\gamma}_j(n)},\cdots)\rightarrow\varrho=(\varrho_1,\cdots,\varrho_j,\cdots)$. We also define $\varrho_+$ as the limiting $l_1$-norm: $\varrho_+=\sum_j\varrho_j$. In addition, we assume that \textbf{(e)} $\alpha^2/\sigma_1^2\leq c$ for a global constant $c>0$, where $\alpha= \twonorm{\bm{\gamma}_1}=\twonorm{\Gamma}$ with $\Gamma=(\bm{\gamma}_1,\cdots, \bm{\gamma}_l)$

\end{assumption}

The following assumption is an extension of the assumption employed in \cite{chun2010sparse} to the tensor coefficient for the characterization of the PLS model. The tensor PLS uses the first loading vectors of $S_{X\tY_k}$ across $k\in [K]$. Hence, we impose a low rank structure on the covariance tensor $\Sigma_{X\tY}$ and provide the derivation of the asymptotic results.

%Hence, the remaining bases of the space spanned by the first mode of $S_{X\tY}$ hinders the characterization of the asymptotic properties of tensor partial least squares. Once we assume that $\tB=\vartheta (\bm{u}_1\circ \bm{u}_2\circ \bm{u}_3)$, we have $\Sigma_{X\tY}=\tB\times_1 \Sigma_{X}=\tilde{\vartheta} (\Sigma_{X}\bm{u}_1/\twonorm{\Sigma_{X}\bm{u}_1}\circ \bm{u}_2\circ \bm{u}_3)$, and thus $\Sigma_{X\tY}$ becomes rank 1, and it enables the derivation of the asymptotic results.

\begin{assumption}\label{asm2}
Assume that each entry of error tensor $\tF$ follows $i.i.d$ $N(0,1)$. Assume also that the Covariance tensor $\Sigma_{X\tY}$ is orthogonally decomposable (odeco), i.e., $\Sigma_{X\tY}=\sum_{r=1}^R\vartheta_r (\bm{u}_{1r}\circ \bm{u}_{2r}\circ \bm{u}_{3r})$ with $R\leq K$, where $\bm{u}_{mr}\in \mathbb{S}^{d_m-1}$, $\bm{u}_{mr}^T\bm{u}_{ms}=0$ for all $r\neq s$ and modes $m=1,2,3$ with $\mathbb{S}^{p-1}$ defined as a unit sphere on $\mathbb{R}^p$ with $\vartheta_1\geq\cdots\geq\vartheta_R>0$. %In addition, for $\delta= \min\{\vartheta_1-\vartheta_2,\vartheta_2-\vartheta_3,\cdots, \vartheta_R\}$ with $\vartheta_1>\vartheta_2>\cdots>\vartheta_R>0$, we have $\delta\geq C(d_1d_2d_3)^{1/4}/\sqrt{n}$ for a global constant $C$.
\end{assumption}
The odeco assumption is a way to extend the Singular Value Decomposition (SVD) for matrices to higher order tensors. In addition, while \cite{chun2010sparse} considers only rank 1 covariance matrix in the corresponding assumption, we generalize this to rank $R$.

We next employ a generalized version of the so called \cite{helland1994comparison} condition (as in  \cite{chun2010sparse} for the proof of matrix PLS consistency). This condition implies that each first mode loading of the covariance tensor $\Sigma_{X\tY}$ is contained in the space spanned by a set of eigenvectors of the covariate covariance matrix $\Sigma_{X}$.

First note that we can restate Assumption~\ref{asm1} as $X=V\Gamma^T+\sigma_1E$ for $\Gamma=(\bm{\gamma}_1,\cdots,\bm{\gamma}_l)\in \mathbb{R}^{d_1\times l}$, $V=(\bm{v}_1\cdots,\bm{v}_l)\in \mathbb{R}^{n\times l}$, $E=(\bm{e}_1^T\cdots,\bm{e}_n^T)^T\in\mathbb{R}^{n\times d_1}$, where $\bm{v}_j=(v_j^1,\cdots,v_j^n)$. Then, we have $\Sigma_{X}=\Gamma\Gamma^T+\sigma_1^2I$.

\begin{condition} \label{cdtn1}
Let $\Gamma_r\in\mathbb{R}^{d_1\times K_r}$ be a $\mathcal{S}_r\subset [K]$ column extraction of $\Gamma\in \mathbb{R}^{d_1\times l}$ for each $r\in [R]$ and let $\mathcal{S}_r$ be disjoint each other with $K=\sum_{r=1}^RK_r$($\leq l$) and  $K_r=|\tS_r|$. We have $\bm{u}_{1r}\in \mathcal{C}(\Gamma_r)$ for each $r\in [R]$, where $\mathcal{C}(A)$ is defined as the column of a matrix $A$. 
\end{condition}

Here, the Condition \ref{cdtn1} is equivalent to Condition 1 imposed in \cite{chun2010sparse}, when $d_2=d_3=1$ and equivalent to Condition 2 in \cite{chun2010sparse}, when $R=1,d_3=1$ for $\Sigma_{X\tY}$.

Under these two assumptions and the condition, which are higher order generalizations of those imposed in \cite{chun2010sparse}, we can subsequently prove that $\Sigma_{X\tY_k}=
\Sigma_{X\tY}-\sum_{j=1}^{k-1}\Sigma_{X\tY_{j}}\times_1\left(\Sigma_{X_j}\bm{q}_j^{(1)}(\bm{q}_j^{(1)T}\Sigma_{X_j}\bm{q}_j^{(1)})^{-1}\bm{q}_j^{(1)T}\right)$ is an odeco tensor for each $k\geq 2$, where $\Sigma_{X_k}=\Sigma_{X}-\sum_{j=1}^{k-1}\Sigma_{X_j}\bm{q}_j^{(1)}(\bm{q}_j^{(1)T}\Sigma_{X_j}\bm{q}_j^{(1)})^{-1}\bm{q}_j^{(1)T}\Sigma_{X_j}$, $\Sigma_{X\tY_1}=\Sigma_{X\tY}$, $\Sigma_{X_1}=\Sigma_{X}$ and $\bm{q}_j^{(1)}$ is the population level first loading vector of the rank$-(1,1,1)$ decomposition of $\Sigma_{X\tY_j}$. Then, by employing the perturbation bound results for odeco tensors of \cite{auddy2023perturbation}, we can obtain correct estimation on the loading matrices $W,Q_2,Q_3$. Consequently, the PLS estimator $\hat{\tB}_{PLS}$  converges with the rate $O_p\left(K\sqrt{(d_1+d_2+d_3)/n}\right)$ to the target parameter $\tB$. We now state the convergence result of $\hat{\tB}_{PLS}$ as described.

\begin{thm}\label{thm1}
 Under Assumption \ref{asm1}, Assumption \ref{asm2}, Condition \ref{cdtn1}, and an initialization assumption for Lemma \cref{supp-lem_Sigma_error}, for a fixed $K$, the Partial Least Squares coefficient tensor estimator $\hat{\tB}_{PLS}$ satisfies $\Fnorm{\hat{\tB}_{PLS}-\tB}\leq O_p\left(K\sqrt{\frac{d_1+d_2+d_3}{n}}\right)$, as $(d_1+d_2+d_3)/n\rightarrow 0$.
\end{thm}

Theorem \ref{thm1} emphasizes that the convergence of the tensor partial least squares coefficients to the true coefficients cannot be guaranteed unless $(d_1+d_2+d_3)/n\rightarrow 0$. Moreover, the fact that each PLS component $X\bm{w}_k$ is a linear combination of the entire $d_1$ variables challenges the interpretation of these PLS components, especially when $X$ is high-dimensional. In fact, such a scenario arises in many fields, including genomics, social networks, and neuroscience. To accommodate this critical issue that arise in applications,  we introduce a sparse higher order partial least squares model and show that coefficient tensor estimator in this model attains desirable asymptotic properties even with the high-dimensional regime in the next section.

\section{Sparse higher order partial least squares}

\subsection{Sparse higher order partial least squares model}

\begin{figure}[!ht]\centering
    % \hspace{-3cm}
   \includegraphics[width=1.05\textwidth]{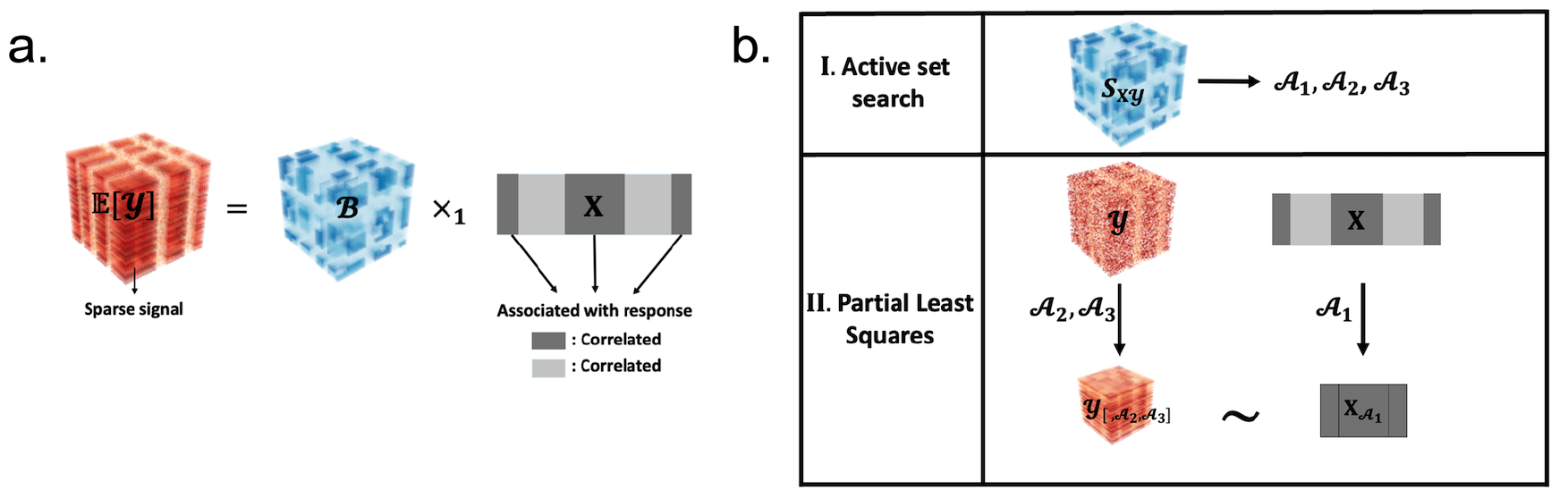}
\caption{SHOPS overview.  \textbf{a.} Problem formulation of the SHOPS model.  SHOPS considers cases where the covariates/predictors in $X$ are highly correlated, and only a relatively small subset of the these associates with the mean tensor $\mathbb{E}[\tY]$. This sparsity  is captured by sparse first mode of the coefficient tensor $\tB$, where darker colors represent higher coefficient magnitudes. In addition, vast majority of the second and third mode entries of the mean tensor $\mathbb{E}[\tY]$ are zeros, as represented by the sparsity of the second and third modes of the coefficient tensor $\tB$. SHOPS can simultaneously select important variables (darker shades in $X$) and detect mean tensor entries due to pure noise, while performing dimensionality reduction for the highly correlated variables.  \textbf{b.} Summary of the SHOPS algorithm. SHOPS first identifies active sets $\tA_{mk}$ with an active set finding algorithm that employs soft-thresholding and hard-thresholding (Step \textbf{I}). Then, it fits ordinary tensor HOPLS onto reduced $\tY$ and $X$ to estimate non-zero entries of the coefficient tensor $\tB$ (Step \textbf{II}) in an iterative manner across $k\in [K]$ with successive deflation on $S_{X\tY}$.
} \label{fig:intro} 
\end{figure}

We start out this section by formally defining the Sparse Higher Order Partial least Squares (SHOPS) model and present the algorithm for estimating the SHOPS model parameters in the next section. The SHOPS model for $\tY\in \mathbb{R}^{n\times d_2\times d_3}$ and $X\in \mathbb{R}^{n\times d_1}$ is
          \begin{align}\label{model}
          \tY&=\tB\times_1X +\sigma_2\tF\notag\\
          &=\tB_W\times_1XW +\sigma_2\tF,
          \end{align}
where we have
\begin{equation*}
\tB_W=\mathcal{G}\times_1(W^T\Sigma_{X}W)^{-1}\times_2 Q_{2}\times_3 Q_{3}    
\end{equation*}
with $W=[\bm{w}_1\cdots,\bm{w}_K]\in \mathbb{R}^{d_1\times K}$, $Q_2=[\bm{q}^{(2)}_1\cdots,\bm{q}^{(2)}_K]\in \mathbb{R}^{d_2\times K}$, $Q_3=[\bm{q}^{(3)}_1\cdots,\bm{q}^{(3)}_K]\in \mathbb{R}^{d_3\times K}$, and 
$\mathcal{G}=\mbox{superdiag}\left(g_k
\right)_{k=1}^K\in\mathbb{R}^{K\times K\times K}$, where $g_k= \Sigma_{X\tY}
\times_1 \bm{w}_k^T\times_2 \bm{q}_k^{(2)T}\times_3 \bm{q}_k^{(3)T}$. Now, the sparse latent direction vectors $\bm{w}_{k},\bm{q}_k^{(2)},\bm{q}_k^{(3)}$ are the successive loading solutions of 
\begin{equation*}
\max\limits_{\bm{s}_1,\bm{s}_2,\bm{s}_3}{ |\Sigma_{X\mathcal{Y}}\times_1 \bm{s}
_1\times_2 \bm{s}_2\times_3 \bm{s}_3|^2 },    
\end{equation*}
subject to $\bm{s}_1^T\Sigma_{X}\bm{w}_j=0$  for $ j\in [k-1]$, and $\bm{s}_1^T(I_{d_1}-W_{k-1}W^+_{k-1})\bm{s}_1=1$, $\bm{s}_2\in\mathbb{S}^{d_2-1}$, $\bm{s}_3\in\mathbb{S}^{d_3-1}$, $\zeronorm{\bm{s}_m}\leq s_m$ for $m=1,2,3$. Note that $s_m= |\tA_m^*|$, where $\tA_m^*=\cup_{k=1}^K{\tA_{mk}^*}$ with $\tA_{mk}^*\subset [d_m]$ defined as active sets, i.e., non-zero entries, of $\bm{w}_k$, $\bm{q}_k^{(2)}$,$\bm{q}_k^{(3)}$ for modes $m=1,2,3$ and factor $k\in [K]$, respectively.

\subsection{Implementation of the SHOPS model and the corresponding algorithms}

Following a similar approach to the SPLS algorithm of \cite{chun2010sparse} and \cite{chung2010sparse}, 
the main  SHOPS  algorithm estimates the active sets $\tA_{mk}^*$ for each $k\in [K]$ and for $ m\in [3]$ and then implements the reduced tensor partial least squares regression algorithm on the estimated supports $\hat{\tA}_{m}^k= \bigcup_{j=1}^k\hat{\tA}_{mj}$ with $k$ latent factors as described in Section~\ref{subsec:TPLSmodel}  (\textbf{Fig.}~\ref{fig:intro}b). The specific steps of the algorithm are presented in Algorithm~\ref{alg:SHOPS}.

\begin{algorithm}\footnotesize
\caption{Sparse Higher Order Partial least Squares (SHOPS)}\label{alg:SHOPS}
\begin{algorithmic}

\State \textbf{Initialize} ${\tY}_1=\tY$, ${X}_1=X$, $\hat{\mathcal{B}}_{SHOPS}=\bm{0}\in\mathbb{R}^{d_1\times d_2\times d_3}$, $\hat{\tA}_m^0=\emptyset$.

\For {$k=1,\cdots, K$}.
%\State 1. Calculate $S_{X\tY_k}={\tY}_k\times_1{X}_k/n$.
\State 1. Obtain $\hat{\mathcal{A}}_{mk}\leftarrow \mbox{Activeset}(\tY_k,X_k)$  for $m\in [3]$ via Algorithm \ref{alg:activeset}.

\State 2. Update $\hat{\tA}_m^k=\hat{\tA}_{mk}\cup \hat{\tA}_m^{k-1}$ for $m\in [3]$.

\State 3. Fit $k$ component HOPLS via Algorithm \ref{alg:HOPLS}
with $\mathcal{Y}_{[\, , \hat{\mathcal{A}}^k_{2},\hat{\mathcal{A}}^k_{3}]}$, $X_{\hat{\mathcal{A}}^k_{1}}$ to obtain $\tilde{\tB}_k\in\mathbb{R}^{\bigtimes_{m=1}^3{|\hat{\tA}^k_m|}}$ and reduced PLS direction $\tilde{W}_k\in \mathbb{R}^{|\hat{\tA}^k_1|\times k}$.

\State 4. Update $\hat{B}_{SHOPS}$ with $\tilde{\tB}_k$ through non-zero entries $\hat{\tA}^k_m$, for $m\in [3]$. 

\State 5. Update PLS direction $\hat{W}_{k}$ with $\tilde{W}_k$ through non-zero entries $\hat{\tA}^k_1$.

\State 6. Update ${\tY}_{k+1}=\tY\times_1(I-X\hat{W}_k(\hat{W}_k^TS_X\hat{W}_k)^{-1}\hat{W}_k^TX^T)$.

\EndFor

\textbf{Return} $\hat{B}_{SHOPS}$

\end{algorithmic}
\end{algorithm}

At the final $K$th step, this algorithm yields the estimate $\hat{\tB}_{SHOPS}$,  $\hat{\tA}_m^K$ entries of which are estimated through $K$ component HOPLS implemented only on the active entries $\hat{\tA}_m^K$ for $ m\in [3]$, while the inactive entries, i.e. $\hat{\tA}_m^{KC}= [d_m]/\hat{\tA}_m^{K}$, are all left as zero.      
For the active set finding step  in the SHOPS main algorithm (step 1  in Algorithm~\ref{alg:SHOPS}), we suggest a two-step thresholding scheme to find the non-zero entries of the covariance tensor $\Sigma_{X\tY_k}$ (Algorithm~\ref{alg:activeset}). This algorithm operates by first soft-thresholding the sample coefficient tensor entry-wise and then hard-thresholding the loading vectors of the soft-thresholded tensor. A similar  two-step thresholding algorithm was also employed in \cite{deshpande2014sparse} for Sparse Principal Component Analysis (SPCA) and \cite{gao2017sparse} for Spasrse Canonical Correlation Analysis (SCCA). 
A sample splitting procedure is also employed within the algorithms. The main benefit of the sample splitting is the  decoupling of  the dimension reduction (HOOI) and the selection of active sets (hard-thresholding) \citep{cai_sparse_2012,deshpande2014sparse,gao2017sparse}.

\begin{algorithm}[!ht]\footnotesize
\caption{Active set finding algorithm}\label{alg:activeset}
\begin{algorithmic}

\Require $\tY_k$: response tensor, $X_k$: covariate matrix , $\tau_k$, $T_{mk}$

\State 1. Compute $S_{X\tY_k}=\tY_k[\Omega_1 , \, ,\,]\times_1 X_k[\Omega_1 , \,]^T/n_1$, $S_{X\tY_k}^\prime=\tY_k[\Omega_2 , \, ,\,]\times_1 X_k[\Omega_2 , \,]^T/n_2$, with $n_1=n_2=n/2$, and with $\Omega_1$: $n_1$ random samples from $[n]$, $\Omega_2$: $n_2$ samples $[n]/\Omega_1$.

\State 2. Compute the soft-thresholded covariance tensor $\eta(S_{X\tY_k})$ :
          \begin{equation*}
          \eta(S_{X\tY_k})_{i_1i_2i_3}=\begin{cases}
          (S_{X\tY_k})_{i_1i_2i_3}-\frac{\tau_k}{\sqrt{n_1}}\mbox{ if }(S_{X\tY_k})_{i_1i_2i_3}\geq \frac{\tau_k}{\sqrt{n_1}}\\
          0\qquad\qquad\qquad\mbox{   if }|(S_{X\tY_k})_{i_1i_2i_3}|<\frac{\tau_k}{\sqrt{n_1}}\\
          (S_{X\tY_k})_{i_1i_2i_3}+\frac{\tau_k}{\sqrt{n_1}}
          \mbox{ if }(S_{X\tY_k})_{i_1i_2i_3}\leq -\frac{\tau_k}{\sqrt{n_1}}.
          \end{cases}
          \end{equation*}

\State 3. Let $\hat{\bm{q}}^{(m)}_{k0}$ be the $m$th-mode first loading vector of $\eta(S_{X\tY_k})$ obtained from rank-(1,1,1)
$\mbox{HOOI}(S_{X\tY_k})$

\State 4. Obtain $\hat{\tA}_{mk}\coloneqq \{i\in [d_m]: \left|\left({S_{X\tY_k}}_{(m)}^\prime\hat{\bm{q}}^{-(m)}_{k0}\right)_i\right|
\geq T_{mk}\}$ for $m=1,2,3$, where $\hat{\bm{q}}^{-(m)}_{k0}=\bigotimes\limits_{l\neq m}\hat{\bm{q}}^{(m)}_{k0}.$

\State \textbf{Return} active sets $\hat{\tA}_{1k},\hat{\tA}_{2k},\hat{\tA}_{3k}$.
          
\end{algorithmic}
\end{algorithm}

In practice,  we employ an approach similar to \cite{deshpande2014sparse} for setting the soft-thresholding parameter $\tau_k$ and a similar approach to that of \cite{chun2010sparse} for setting the hard-thresholding parameter $T_{mk}$ (Section~\ref{thrchoice}). This results in choosing only one parameter ($0<\nu<1$), which corresponds to the fraction of maximum magnitude of the loading estimate $\left|\left({S_{X\tY_k}}_{(m)}^\prime\hat{\bm{q}}^{-(m)}_{k0}\right)_i\right|$. 
In the next section, we investigate the probabilistic performance of the active set finding algorithm and derive the asymptotic results of the SHOPS estimator.

\subsection{Asymptotic results for sparse higher order partial least squares}

In this section, we derive the asymptotic properties  of the SHOPS estimator when the dimension of the entire coefficient tensor ($\sum_{m=1}^3d_m$) grows possibly faster than the sample size $n$. 
We first assume that, as in the ordinary tensor partial least squares, Assumption~\ref{asm1} holds. In addition, we assume a sparse modification of Assumption~\ref{asm2}.

\begin{assumption}\label{asm3}
Assume that each entry of error tensor $\tF$ follows $i.i.d$ $N(0,1)$. Assume also that the Covariance tensor $\Sigma_{X\tY}$ is odeco, and for $\Sigma_{X\tY}=\sum_{r=1}^R\vartheta_r (\bm{u}_{1r}\circ \bm{u}_{2r}\circ \bm{u}_{3r})$ such that $\bm{u}_{mr}\in \mathbb{S}^{d_m-1}$ for $ r\in[R]$, $\vartheta_1\geq\cdots\geq\vartheta_R>0$, and $\zeronorm{U_m}=  s_m$, for $m=1,2,3$, where $U_m=[\bm{u}_1,\cdots, \bm{u}_R]$.
\end{assumption}

Lastly, we impose the sparse form of the \cite{helland1994comparison} condition similar to Condition \ref{cdtn1} and to that of the sparsity assumption on loading matrix in envelope-based SPLS of \cite{zhu2020envelope}. Once we impose Assumption~\ref{asm3} on Condition~\ref{cdtn2} stated below, we can ensure that the active (non-zero) rows of $W,Q_2,Q_3$ are the same as those of $U_1,U_2,U_3$ respectively, which facilitates the theoretic argument of the asymptotic properties.

\begin{condition} \label{cdtn2}
Let $\Gamma_r\in\mathbb{R}^{p\times K_r}$ be a $\mathcal{S}_r\subset [K]$ column extraction of $\Gamma\in \mathbb{R}^{p\times l}$ for each $r\in [R]$ and let $\mathcal{S}_r$ be disjoint of each other with $K=\sum_{r=1}^RK_r$($\leq l$) and  $K_r=|\tS_r|$. We have $\bm{u}_{1r}\in \mathcal{C}(\Gamma_r)$ for each $r\in [R]$. It also holds that $\Gamma_{r\tA_1^{*C}}=\bm{0}\in\mathbb{R}^{(d_1-s_1)\times K},
$ where $\Gamma_{r\tA_1^{*C}}$ is the $\tA_1^{*C}$ rows of $\Gamma_{r}$ for $ r\in [R]$.
\end{condition}

For the consistency proof of SHOPS, we first show that the active set finding algorithm correctly finds the true active entries $\tA_1^*,\tA_2^*,\tA_3^*$ with high probability after $K$th SHOPS iteration (Lemma~\cref{supp-lem:generalk}), and prove the consistency of $\hat{\tB}_{SHOPS}$ given true active sets $\tA_1^*,\tA_2^*,\tA_3^*$. Then, by combining the two results, we obtain the consistency of $\hat{\tB}_{SHOPS}$ (Theorem~\ref{thm2}).
          
Specifically, we argue that the active set finding algorithm identifies true non-zero entries with high probability across $k\in [K]$ with the following three steps: \textbf{Step 1:} Soft-thresholded covariance tensor $\eta(S_{X\tY_k})$ is close to the true covariance tensor $\Sigma_{X\tY_k}$ with high probability. \textbf{Step 2:} The loading vector estimates of the soft-thresholded covariance tensor $\eta(S_{X\tY_k})$ are close enough to the true loading vectors of $\Sigma_{X\tY_k}$. \textbf{Step 3:} The hard-thresholding step correctly recovers the non-zero entries $\tA_{mk}^*$ with high probability. 
Lemmas~\cref{supp-lem:soft}, \cref{supp-lem:dist}, and \cref{supp-lem:activeset} in the  Section~\ref{supp-sectionS:proofs} prove  \textbf{Step 1}, \textbf{Step 2}, and \textbf{Step 3}, respectively, for $k=1$, and Lemma~\cref{supp-lem:generalk} proves these for general $k\in [K]$. With these results, we are able to establish that the active set finding algorithm correctly finds the true active sets $\tA_m^*=\bigcup_{k=1}^K\tA_{mk}^*, \mbox{ for } m\in [3]$ with high probability as the sample size grows faster than the number of non-zero entries, and consequently, we can establish that the tensor SHOPS estimator $\hat{\tB}_{SHOPS}$ achieves $\sqrt{n}$-consistency to $\tB$ under the high-dimensional regime as follows.

\begin{thm}\label{thm2}
Assume that Assumptions~\ref{asm1}, \ref{asm3} and Condition~\ref{cdtn2} hold, and that $\vartheta_{(1)k}-\vartheta_{(2)k}\geq C(\alpha+\sigma_1)\{(\alpha+\sigma_1)\beta+\sigma_2\}\sqrt{s_1s_2s_3\log{(d_1d_2d_3)}/n}$ %and  $\vartheta_{(1)k}/\vartheta_{\min k}\leq C$
holds for $ k\in [K]$ for a global constant $C$. Further assume that 
\begin{equation}\label{eq:signaltonoise}
\mbox{(Signal to noise ratio): }\vartheta_*\geq C_0{(\alpha+\sigma_1)\{(\alpha+\sigma_1)\beta+\sigma_2\}}\sqrt{\frac{s_1s_2s_3\log(d_1d_2d_3)}{n}},
\end{equation}
and that  there exists $\lambda_{mk},\gamma>0$ satisfying 
\begin{equation}\label{eq:seperability}
\mbox{(Separability): }|u_{m(1)ki}|\geq \frac{\lambda_{mk}}{\sqrt{s_1s_2s_3}}    
\end{equation}
and $|u_{m(r)ki}/u_{m(1)ki}|\leq \gamma$ for all $k\in [K]$, $i\in [\tA_{mk}^*]$, $r\geq 2$, where $\beta= \Fnorm{\tB}$, $\Sigma_{X\tY_k}=\sum_{r=1}^R\vartheta_{(r)k}(\bm{u}_{1(r)k}\circ \bm{u}_{2(r)k}\circ \bm{u}_{3(r)k})$, $\vartheta_{(1)k}\geq \cdots \geq\vartheta_{(R)k} \geq  0$, and $\vartheta_*=\min\limits_{k\in [K]}\vartheta_{(1)k}$. Lastly, assume that 
\begin{equation}\label{crosscov}
\mbox{(Identifiability): }\max_{i\in\tA_1^{*}}\twonorm{\mbox{Cov}(X_i,X_{\tA_1^{*C}})}\leq C_0\sigma_1^2 \sqrt{\frac{\log{d_1}}{\sum_{m=1}^3{s_m}}},   
\end{equation}
for a constant $C_0>0$. Then, with the choice of soft thresholding parameters $\tau_k>C^*_k(\alpha+\sigma_1)\{(\alpha+\sigma_1)\beta+\sigma_2\}\sqrt{\log(d_1d_2d_3)}$ and $T_{mk}=\frac{\vartheta_*\lambda_{mk}}{2\sqrt{s_1s_2s_3}}$ for Algorithm~\ref{alg:activeset} for constants $C^*_k>0$, for $ m\in [3]$ and $k\in [K]$, as $n,d_1,d_2,d_3\rightarrow \infty$, we have
\begin{equation*}
\textbf{(i) }\mathbb{P}\left(\hat{\tA}_m^K=\tA_m^*\right)\geq 1-o(1) \mbox{ and }\textbf{(ii) }\Fnorm{\hat{\tB}_{SHOPS}-\tB}\leq O_p\left(K\sqrt{\frac{s_1+s_2+s_3}{n}}
\right).
\end{equation*}

\end{thm}

Theorem~\ref{thm2} implies that we are able to find the true active sets $\tA_m^*$ with probability tending to 1. Theorem~\ref{thm2} also demonstrates the clear advantage of using SHOPS model over using ordinary tensor PLS. As emphasized in Theorem~\ref{thm1}, the ordinary tensor PLS could be inconsistent when $\sum_{m=1}^3d_m\gtrsim  n$. In contrast, SHOPS model can still accurately estimate $\tB$ with the rate of $O_p\left(\sqrt{\sum_{m=1}^3s_m/n}\right)$. In scientific applications such as the single cell 3D genome organization with  scHi-C and single cell DNA methylation data, we expect  methylation of a small number of genomic regions ($s_1$) to significantly affect the long-range physical contact between a small number of genomic locus pairs ($s_2$, $s_3$). Therefore, we can reasonably expect to obtain markedly better accuracy of the coefficient tensor estimator. 

The key advantage of utilizing PLS framework in tensor regression over such methods is in its ability to handle multicollinearity issues existing in high-dimensional covariates, which will be more clearly demonstrated in the simulation studies presented in Section~\ref{sec:simulation}, even if the rate of estimation accuracy is similar to that of general low rank and sparse tensor regressions as described in \cite{raskutti2019convex}.

We note that the signal to noise ratio assumption \eqref{eq:signaltonoise} with a multiplicative order of $s_m$ is similarly required by \cite{gao2017sparse} in the context of SCCA for deceomposing $S_{XY}$ and also by \cite{zhang2019optimal} in the sparse tensor decomposition context. In addition, the separability assumption \eqref{eq:seperability} on the active loading entries is similarly imposed in support recovery literature \citep{meinshausen2006high,deshpande2014sparse}. The additional assumption on the bounded covariance between the active variables and in-active variables in \eqref{crosscov} is required specifically by the fact that the sample version of the projection matrix was utilized to obtain $S_{X\tY_k}$ for each $k\in [K]$, and it is closely related to how detectable the active variables are compared to the inactive variables. In fact, the simulation studies investigated by SPLS methods \citep{chun2010sparse,chung2010sparse} are with block covariance structures, i.e., with $\max_{i\in\tA_1^{*}}\twonorm{\mbox{Cov}(X_i,X_{\tA_1^{*C}})}=0$, which is a special case of \eqref{crosscov}.

Finally, it is worth noting that when $d_3=1$ (or when $d_2=d_3=1$), the Algorithm~\ref{alg:SHOPS} reduces to the multivariate (or univariate) SPLS algorithm in \cite{chun2010sparse},  with the only difference being the choice of the active set finding algorithm in Step 2, as the HOOI reduces to the rank-$1$ SVD of $\Sigma_{XY}$. Then, Theorem~\ref{thm2} reveals the previously unknown asymptotic behavior of the SPLS estimator $\hat{B}_{SPLS}$, which appears to enjoy wide popularity in diverse set of applications.

\section{Practical considerations for selecting the thresholding parameter and the number of components}\label{thrchoice}

While there are several parameters $\tau_k$, $\{T_{mk}|m\in [3]\}$ to set for the thresholding operations in Algorithm~\cref{alg:activeset}, we suggest a way to set the soft thresholding parameter $\tau_k$ borrowing an argument similar to the one provided in \cite{deshpande2014sparse}. As for $T_{mk}$, we suggest a way similar to \cite{chun2010sparse} for choosing single parameter $\nu$, which corresponds to a fraction of the maximum component values.

\subsection{Choice of the soft-thresholding parameter $\tau_k$} 
Consider the null hypothesis that $\mathbb{E}({S_{X\tY}})_{ijk}=0$, and the standard deviation of (the signal part of) ${S_{X\tY}}_{ijk}$, $\sigma({S_{X\tY}}_{ijk})\approx\sigma_2\sqrt{(\twonorm{\bm{\gamma}^i}^2+\sigma_1^2)}/\sqrt{n_1}$ obtained from the SHOPS model \ref{model}, where $\bm{\gamma}^i\in \mathbb{R}^{l}$ is $i$th row of the matrix $\Gamma$ in Assumption \ref{asm1}. We ``reject'' the null if $|{S_{X\tY}}_{ijk}|>\nu_0\sigma({S_{X\tY}}_{ijk})$ for some $\nu_0\geq 0$. Note that for the $i$th covariate, we have $\sqrt{{(\Sigma_{X})}_{ii}}=\sqrt{(\twonorm{\bm{\gamma}^i}^2+\sigma_1^2)}$ by Assumption \ref{asm1}. Therefore, we estimate the term $\sqrt{(\twonorm{\bm{\gamma}^i}^2+\sigma_1^2)}$ as the standard deviation of $X_i$, e.g., $\mbox{MAD}(X_i)/\Phi^{-1}(3/4)$ or $\mbox{s.d.}(X_i)$, where $\mbox{s.d.}(\cdot)$ refers to standard deviation,  $\mbox{MAD}(\cdot)$ is the median absolute deviation and $\Phi(\cdot)$ is the c.d.f of standard normal distribution. As for the estimate of $\sigma_2$ in the model, we suggest using the sample standard devation estimate or $\mbox{MAD}(\tY)/\Phi^{-1}(3/4)$ or $\mbox{s.d.}(\tY)$. As a result, we recommend to use the estimate
\begin{equation*}
\hat{\sigma}({S_{X\tY}}_{ijk})=\frac{1}{\sqrt{n_1}}\frac{\mbox{median}_{i\in [d_1]}(\mbox{MAD}(X_i))}{\Phi^{-1}(3/4)}\frac{\mbox{MAD}(\tY)}{\Phi^{-1}(3/4)} \quad \left(\mbox{or }\frac{1}{\sqrt{n_1}}\mbox{median}_{i\in [d_1]}(\mbox{s.d.}(X_i))\mbox{s.d.}(\tY)\right),    
\end{equation*}
i.e., set the thresholding parameter $\tau_1 \equiv \nu_0\sqrt{n_1}\hat{\sigma}({S_{X\tY}}_{ijk})$, and perform soft-thresholding based on the criterion $|{S_{X\tY}}_{ijk}|>\tau_1/\sqrt{n_1}$. Similarly, we can obtain $\tau_k$ by setting $\tau_k \equiv \nu_0\sqrt{n_1}\hat{\sigma}({S_{X\tY_k}}_{ijk})$, where we replace $\tY$ by $\tY_k$ for the standard deviation estimate calculation for general $k\in [K]$. We suggest to set $\nu_0=\Phi^{-1}(0.95)\approx 1.65$.

\subsection{Choice of the hard-thresholding parameter $T_{mk}$} 

Note that we can reformulate the step 4 of Algorithm~\ref{alg:activeset} as 
\begin{equation*}
\hat{\tA}_{mk}=\left\{i\in [d_m]|\left|\left({S_{X\tY_k}^\prime}_{(m)}\bm{q}_{k0}^{-(m)}\right)_i\right|>\nu \max_{j\in [d_m]}\left|\left({S_{X\tY_k}^\prime}_{(m)}\bm{q}_{k0}^{-(m)}\right)_j\right|\right\},    
\end{equation*}
for a $0<\nu<1$. Note that $\nu$ plays the role of hard thresholding parameter and we retain the entries, whose component values are greater than some fraction of the maximum value. Similar approach is applied in \cite{friedman2004gradient} and \cite{chun2010sparse}. This enables tuning the hard thresholing parameter within a bounded range across $m\in [3]$. In practice, $\nu$ can be tuned by cross-validation (CV) along with the number of latent components $K$ of the SHOPS. A single $\nu$ is tuned for all $K$, since tuning multiple parameters is computationally expensive and could produce non-unique optimizers for the CV \citep{chun2010sparse}. %When the  explanatory variables are low dimensional  (small $d_1$) and the sparsity on the first mode of the coefficient tensor $\tB$ is not of interest, one can simply set $T_{1k}=0$ and perform CV by only varying the thresholding parameters of the second and third modes. Same strategy can be used for the second and third modes.
When CV for choosing $\nu$ is prohibitive due to computational complexity, we suggest a Gaussian Mixture Model based approach to determine thresholding parameters in Supplementary Section \cref{supp-GMM_thr}.

\subsection{Choice of number of latent components}\label{latent_factors}   
Conditions 1 and 2 imply that the number of eigenvectors of $\Sigma_{X}$, denoted as $l$, cannot be smaller than the latent PLS directions $K$. Hence, $l$ is a natural upper bound when performing  CV for $K$.  We recommend performing eigen-decomposition of $S_{X}$ and choosing the estimate of $l$ as the elbow point of the eigenvalue scree plot. Then, we perform a CV with the elbow point $K_{\max}$ as the upper bound for it. In fact, by the way Algorithm~\ref{alg:SHOPS} is implemented, we do not need to fit SHOPS for each $k\in [K_{\max}]$, since a single SHOPS run with $K_{\max}$ component already yields coefficient tensor estimates for $\forall k\leq K_{\max}$.

\section{Simulation studies}\label{sec:simulation}

We next provide a wide range of simulation settings to empirically evaluate the properties of SHOPS and benchmark it against alternatives.  The overall data generating schemes of these simulations were set as follows :

\noindent \textbf{Covariate matrix:} With independent hidden variables $\bm{h}_j\sim N(0,\lambda^2 I_n)$ for each $j\in [K+1]$, we set $X=(\bm{x}_1,\cdots,\bm{x}_p)$ as
\begin{equation*}
\bm{x}_i=
\begin{cases}
&a_1\bm{h}_j-\sum_{k\neq j}a_0\bm{h}_k+\bm{\epsilon}_i,\mbox{ if } 1\leq i\leq s,\\
&a_2\bm{h}_j+\bm{\epsilon}_i,\mbox{ if }  s\leq i\leq p,\\
\end{cases}
\end{equation*}
for $p_{j-1}+1\leq i\leq p_j$, where $j\in [K+1]$, $(p_0,\cdots, p_{K+1})=(0,1/Ks,2/Ks,\cdots,s,p)$ and $\epsilon_i$ are drawn from $N(0,I_n)$ independently. Note that for 
\begin{align*}
\bm{\gamma}_j&=(\overbrace{-a_0,\cdots, -a_0}^{(j-1)s/K},\overbrace{a_1,\cdots, a_1}^{s/K},\overbrace{-a_0,\cdots, -a_0}^{s-j/Ks},\overbrace{0,\cdots, 0}^{p-s})\mbox{, for } j\leq K,\\
\bm{\gamma}_{K+1}&=(\overbrace{0,\cdots, 0}^{s},\overbrace{a_2,\cdots, a_2}^{p-s}),
\end{align*}
$a_0,a_1,a_2> 0$ are set to satisfy $\twonorm{\bm{\gamma}_j}=1$ and $\bm{\gamma}_j^T\bm{\gamma}_k=0$ for $j\neq k$. Notice here that the $K+1$ singular values of $X$ are all $\lambda=10$. We considered the number of variables $p=240$ for the simulations, and varied the number of PLS component $K\in \{3,15\}$, depending on the rank of $\tB$ considered, sample size $n\in \{120,360\}$ and the number of active variables $s\in \{30,60\}$.

\noindent \textbf{Coefficient tensor:} We set the coefficient tensor to be orthogonally decomposable:
\begin{align*}
\tB&=\sum_{r=1}^R{\theta_r(\bm{u}_{1r}\circ \bm{u}_{2r} \circ \bm{u}_{3r})},\\&\mbox{ s.t. } \bm{u}_{mr}\in \mathbb{S}^{d_m} \mbox{ and } \bm{u}^T_{mr}\bm{u}_{ms}=0, \mbox{ for } r\in [R],r\neq s,m\in [3],
\end{align*}
where $d_1=p=90$, $d_2=64$, and $d_3=64$. Note first that each first loading vector $\bm{u}_{1r}$ is generated as $\bm{u}_{1r}=\sum_{j=1}^Kc_{jr}\bm{\gamma}_j$, where $c_{jr}\in \{0,1\}$ and $\sum_{j=1}^Kc_{jr}\neq 0$. This is to guarantee that all the $\bm{\gamma}_j$s for $ j\in [K]$ contributes to generate the first loading vectors. Then, by the sparsity pattern of $\bm{\gamma}_j$ for $j\leq K$, the last $p-s$ entries of $\bm{u}_{1r}$ are zeros for $ r\in [R]$, which indicates that the last $p-s$ variables are irrelevant to association between $X$ and $\tY$. Note that we set $K=3$ for $R\in\{1,2\}$ and $K=15$ for $R\in \{9,14\}$.

For a clearer investigation on the performance of each method considered for the simulation in terms of response denoising and active entries recovery, we set the $\theta_r,\bm{u}_{2r},\bm{u}_{3r}$ to make $\tB[1,,]$ %$\left(=\sum_{r=1}^R\theta_r\bm{u}_{1r}[1]\bm{u}_{2r}\circ \bm{u}_{3r}\right)$
visualize one of the visual patterns we considered. These are square ($R=1$), cross ($R=2$), circle ($R=9$), and bat ($R=14$), displayed in the first column of \textbf{Fig.}~\ref{fig:betahats}. The second and third mode entries (white entries) not filled with one of the patterns are set as zeros, which results in sparsity on $\bm{u}_{2r},\bm{u}_{3r}$. Lastly, we set the signal size $\theta_1$ as  $\twonorm{\tB}=5$ for $ R\in \{1,2,9,14\}$.

\noindent\textbf{Response tensor:} We  generated the response tensor based on the model 
\begin{equation*}
\mathcal{Y}=\tB\times_1 X+\sigma_2\tF\in\mathbb{R}^{n\times 64 \times 64},     
\end{equation*}
where $\tF_{ijk}\stackrel{i.i.d}{\sim}N(0,1)$. We considered the noise level $\sigma_2=2$ for the simulations.

We compared SHOPS against methods including SPLS of \cite{chun2010sparse}, Sparse Principal Component Analysis (SPCA) \citep{Erichson2020} regression, HOPLS \citep{zhao2012higher}, envelope tensor response regression method (ENV) of \cite{li2017parsimonious}, Sparse Tensor Response Regression (STORE) of \cite{sun2017store}, and Ordinary Least Squares (OLS). Both  SPCA and OLS are implemented on each vectorized response variable and the multivariate SPLS is applied on the matricized response $\tY_{(1)}$ , whereas HOPLS, ENV, and STORE incorporate the tensor structure of the coefficient tensor. While SPLS and SPCA can perform both dimension reduction and variable selection of $X$, HOPLS, ENV and STORE conduct only dimension reduction. Compared to HOPLS and ENV, STORE can find active entries along the second and third modes of coefficient tensor. For implementation details of these methods, we refer to Supplementary Section~\cref{supp-simul_details}.

Across the settings and methods, we evaluated estimation error $\Fnorm{\hat{\tB}-\tB}$, prediction error $\Fnorm{\tY_{new}-\hat{\tB}\times_1X_{new}}/\sqrt{n}$, where $X_{new},\tY_{new}$ represent test data independent of the training data. In addition, the active set search performance is evaluated by True Positive Rate (TPR) and False Positive Rate (FPR), involving $\mbox{TPR}_1,\mbox{FPR}_1,\mbox{TPR}_{2,3},\mbox{FPR}_{2,3}$, where $\mbox{TPR}_{2,3}=1/2(\mbox{TPR}_2+\mbox{TPR}_3),\mbox{FPR}_{2,3}=1/2(\mbox{FPR}_2+\mbox{FPR}_3)$, $TPR_m=\frac{|\tA_m^{*}\cap \hat{\tA}_m|}{|\tA_m^{*}|}$ and $FPR_m=\frac{|\tA_m^{*C}\cap \hat{\tA}_m|}{|\tA_m^{*C}|}$ for $m\in [3]$. Note that $\hat{\tA}_m$ and $\hat{\tA}_m^C$ denote the estimated active and inactive sets for mode $m$. A total of 30 simulation replicates were carried out for each setting.

\begin{table}[!ht]
\centering
\tiny
\begin{tabular}{c|c|c|c|cc|cc|cc}
\hline
$n$ & $R$ & $s$ & Method & Estimation error & Prediction error & $\mbox{TPR}_1$ & $\mbox{TPR}_{2,3}$ & $\mbox{FPR}_1$ & $\mbox{FPR}_{2,3}$ \\
\hline
\multirow{16}{*}{120} & \multirow{8}{*}{1} & \multirow{4}{*}{30} 
& SHOPS & 0.388 (0.042) & 90.55 (0.090) & 1 (0) & 1 (0) & 0.018 (0.028) & 0.306 (0.097) \\
& & & HOPLS & 0.780 (0.040) & 90.58 (0.090) & 1 (0) & 1 (0) & 1 (0) & 1 (0) \\
& & & SPLS & 1.592 (0.098) & 91.58 (0.097) & 0.997 (0.0121) & 1 (0) & 0.676 (0.142) & 1 (0) \\
& & & SPCA & 2.061 (0.127) & 91.90 (0.102) & 0.997 (0.012) & 1 (0) & 0.465 (0.030) & 1 (0) \\
\cline{3-10}
& & \multirow{4}{*}{60} 
& SHOPS & 0.530 (0.203) & 90.55 (0.090) & 0.995 (0.009) & 1 (0) & 0.0018 (0.0009) & 0.142 (0.084) \\
& & & HOPLS & 0.779 (0.040) & 90.57 (0.091) & 1 (0) & 1 (0) & 1 (0) & 1 (0) \\
& & & SPLS & 1.644 (0.137) & 91.58 (0.093) & 0.990 (0.016) & 1 (0) & 0.741 (0.151) & 1 (0) \\
& & & SPCA & 2.272 (0.111) & 91.90 (0.101) & 0.998 (0.006) & 1 (0) & 0.526 (0.026) & 1 (0) \\
\cline{2-10}
& \multirow{8}{*}{14} & \multirow{4}{*}{30} 
& SHOPS & 2.820 (0.176) & 95.44 (0.562) & 1 (0) & 0.829 (0.121) & 0 (0) & 0.039 (0.047) \\
& & & HOPLS & 2.917 (0.048) & 95.20 (0.189) & 1 (0) & 1 (0) & 1 (0) & 1 (0) \\
& & & SPLS & 3.353 (0.016) & 96.04 (0.113) & 1 (0) & 1 (0) & 0 (0) & 1 (0) \\
& & & SPCA & 3.552 (0.174) & 96.17 (0.110) & 1 (0) & 1 (0) & 1 (0) & 1 (0) \\
\cline{3-10}
& & \multirow{4}{*}{60} 
& SHOPS & 2.727 (0.060) & 95.06 (0.198) & 1 (0) & 0.847 (0.114) & 0 (0) & 0.069 (0.007) \\
& & & HOPLS & 2.924 (0.032) & 95.22 (0.177) & 1 (0) & 1 (0) & 1 (0) & 1 (0) \\
& & & SPLS & 3.392 (0.080) & 96.01 (0.096) & 0.997 (0.007) & 1 (0) & 0 (1) & 1 (0) \\
& & & SPCA & 3.565 (0.020) & 96.16 (0.095) & 1 (0) & 1 (0) & 1 (0) & 1 (0) \\
\hline
\multirow{28}{*}{360} & \multirow{14}{*}{1} & \multirow{7}{*}{30} 
& SHOPS & 0.187 (0.056) & 90.50 (0.038) & 1 (0) & 1 (0) & 0.008 (0.019) & 0.121 (0.081) \\
& & & STORE & 2.031 (0.188) & 90.52 (0.037) & 1 (0) & 1 (0) & 1 (0) & 0 (0) \\
& & & ENV & 6.532 (1.341) & 90.77 (0.29) & 1 (0) & 1 (0) & 1 (0) & 1 (0) \\
& & & HOPLS & 0.415 (0.038) & 90.51 (0.038) & 1 (0) & 1 (0) & 1 (0) & 1 (0) \\
& & & OLS & 132.3 (26.33) & 164.3 (38.38) & 1 (0) & 1 (0) & 1 (0) & 1 (0) \\
& & & SPLS & 0.889 (0.114) & 90.87 (0.041) & 0.997 (0.008) & 1 (0) & 0.759 (0.349) & 1 (0) \\
& & & SPCA & 1.006 (0.019) & 90.99 (0.04) & 1 (0) & 1 (0) & 1 (0) & 1 (0) \\
\cline{3-10}
& & \multirow{7}{*}{60} 
& SHOPS & 0.242 (0.057) & 90.50 (0.038) & 1 (0) & 1 (0) & 0.033 (0.047) & 0.201 (0.103) \\
& & & STORE & 2.047 (0.221) & 90.52 (0.036) & 1 (0) & 1 (0) & 1 (0) & 0.714 (0.065) \\
& & & ENV & 6.634 (1.753) & 90.80 (0.420) & 1 (0) & 1 (0) & 1 (0) & 1 (0) \\
& & & HOPLS & 0.416 (0.043) & 90.51 (0.038) & 1 (0) & 1 (0) & 1 (0) & 1 (0) \\
& & & OLS & 133.7 (34.74) & 166.63 (50.96) & 1 (0) & 1 (0) & 1 (0) & 1 (0) \\
& & & SPLS & 1.335 (1.087) & 90.88 (0.053) & 0.966 (0.069) & 1 (0) & 0.518 (0.266) & 1 (0) \\
& & & SPCA & 1.134 (0.077) & 90.98 (0.043) & 0.998 (0.006) & 1 (0) & 0.805 (0.029) & 1 (0) \\
\cline{2-10}
& \multirow{14}{*}{14} & \multirow{7}{*}{30} 
& SHOPS & 1.685 (0.114) & 92.14 (0.225)& 1 (0) & 0.835 (0.123) & 0 (0)& 0 (0) \\
& & & STORE & 88.76 (442.1) & 216.2 (606.2)  & 1 (0) &  0.788 (0.160) & 1 (0)& 0 (0) \\
& & & ENV & 967.8 (5139) & 1411 (7208)& 1 (0) & 1 (0) & 1 (0)& 1 (0) \\
& & & HOPLS & 1.814 (0.062) & 92.21 (0.113) & 1 (0) & 1 (0) & 1 (0)& 1 (0) \\
& & & OLS & 3903 (20702) & 5466 (29078)& 1 (0) & 1 (0) & 1 (0)& 1 (0) \\
& & & SPLS & 1.904 (0.050) & 92.36 (0.043) & 1 (0) & 1 (0) & 0.579 (0.482)& 1 (0) \\
& & & SPCA & 1.987 (0.017) & 92.46 (0.044) & 1 (0) & 1 (0) & 1 (0)& 1 (0) \\
\cline{3-10}
& & \multirow{7}{*}{60} 
& SHOPS & 1.682 (0.033) & 92.10 (0.096) & 1 (0) & 0.836 (0.122) & 0 (0)& 0.007 (0.016) \\
& & & STORE & 9.887 (9.723) & 107.5 (7.637)& 1 (0) & 0.787 (0.115) & 1 (0)& 0 (0) \\
& & & ENV & 45.77 (88.94) & 116.4 (114.01) & 1 (0) & 1 (0) & 1 (0)& 1 (0) \\
& & & HOPLS & 1.809 (0.032) & 92.20 (0.098) & 1 (0) & 1 (0) & 1 (0)& 1 (0) \\
& & & OLS & 192.5 (374.1) & 252.0 (519.3) & 1 (0) & 1 (0) & 1 (0)& 1 (0) \\
& & & SPLS & 1.978 (0.132) & 92.37 (0.044) & 0.996 (0.008) & 1 (0) & 0.826 (0.282)& 1 (0) \\
& & & SPCA & 1.991 (0.021) & 92.46 (0.046) & 1 (0) & 1 (0) & 1 (0)& 1 (0) \\
\hline
\end{tabular}
\caption{\textbf{Simulation results for $R=1$ and $R=4$}. On each cell, the average value across 30 simulation runs is displayed along with the standard deviation of the values in parentheses. Same results for $R=2$, $R=9$ is in \textbf{Tab.}~\cref{supp-tab:simul_res_R=2_9}.}
\label{tab:simul_res_R=1_14}
\end{table}

\subsection{Comparison with variable selection/dimension reduction methods in terms of prediction power and variable selection}\label{subsec:varselect_dimred}

We compared the performance of SHOPS with respect to variable selection and dimension reduction against other widely used or state-of-the-art methods, including ENV, STORE, HOPLS, SPLS, and SPCA.  All four methods handle dimension reduction for the high dimensional $X$. However, while the first three  incorporate the tensor structure of the response (hence the coefficients $\tB$), they do not perform explicit variable selection. In contrast, SPLS and SPCA enable variable selection, but ignore the tensor structure of the coefficients. 

First, we compare these baseline methods in terms of prediction error to investigate whether the high dimensionality and the correlation structure of $X$ are well incorporated via dimension reduction across methods. Surprisingly, while ENV, STORE are developed to deal with high-dimensional $X$, these return error whenever we have $n<p$ (when $n=120$ in \textbf{Tab.}~\ref{tab:simul_res_R=1_14}), mainly because of the calculation of $S_X^{-1}$. Under this relatively high-dimensional setting ($n<p$), i.e., $n=120$ and $p=240$, SHOPS outperforms HOPLS, SPLS, and SPCA across coefficient rank $R$ and non-sparsity parameter $s$, in general. This is mainly by the fact that SHOPS estimates $\tB$ more accurately than the others, as it can be clearly observed from the estimation errors reported in \textbf{Tab.}~\ref{tab:simul_res_R=1_14}. The performances of SPLS and SPCA are always sub-optimal compared to SHOPS and HOPLS, since these methods cannot incorporate the low dimensional structure of $\tB$ and,  hence, lead to less clear dimension reduction for $X$. %It mainly is from the fact that PLS can distinguish signals coming from $H_K=(\bm{h}_1,\cdots,\bm{h}_K)$ and the irrelevant latent variable $\bm{h}_{K+1}$, while PCA type methods cannot, since the eigenvalues $\lambda_j$ are equal.

When $n>p$, i.e., $n=360$ and $p=240$, SHOPS again yields the lowest prediction error compared to the other methods in general, and the difference is more apparent when $R=14$ compared to that when $R=1$. When the rank $R=1$, HOPLS and SHOPS perform similarly, since with a larger sample size, the estimated PLS coefficient tensor on the in active entries will be close enough to zeros. Nonetheless, the estimation error of SHOPS is still two to four times smaller than that of HOPLS. STORE performs close enough to HOPLS and SHOPS when $R=1$, but performs far worse than those when $R=14$. We note that STORE estimates $\tB$ by acknowledging its low-rank structure; however, the relation between $\bm{u}_{1r}$ and $\bm{\gamma}_j$ is not explicitly handled by the model. In fact, this deficiency of STORE reveals the main benefit of employing PLS framework for tensor regression with highly correlated $X$ over ordinary low-rank (or sparse) tensor regression methods. A similar observation was also made by  \cite{chun2010sparse} and \cite{chung2010sparse} in the context of matrix response regression. Furthermore, STORE is not equipped to estimate $\bm{u}_{1r}$ in a sparse manner, which hinders its  variable selection capabilities. 

Next, we compared the variable selection performances of the methods with evaluation criteria $\mbox{TPR}_1$ and $\mbox{FPR}_1$. Performance evaluation with these criteria directly relates to interpretability of the results as well as more accurate estimation and prediction. Among the methods compared, only SHOPS, SPLS, and SPCA directly handle the variable selection task. For SHOPS, across all the settings, the $\mbox{TPR}_1>0.99$, while the $\mbox{FPR}_1$ is controlled below $0.04$, which represents  successful variable selection by SHOPS. In general, SHOPS and SPLS exhibit a comparable $\mbox{TPR}_1$ against SPCA, with a much lower $\mbox{FPR}_1$. As highlighted earlier, these results also reveal that SPLS-type methods can discern signals from only the important latent variables, while SPCA cannot achieve this, as $\lambda_j=\lambda$ across all $j\in [K+1]$.

\subsection{Comparison with sparse/low-rank tensor regression methods in terms of estimation accuracy and support recovery of the response}\label{subsec:tensor_denoise_recovery}

We next focused on comparing the tensor denoising and active set recovery performances across state-of-the-art methods, specifically for the second and third modes of the coefficient tensor $\tB$. The methods that impose low rankness and sparsity on $\tB$ can perform such tasks, which include SHOPS, HOPLS, ENV, and STORE. Hence, for the comparison, we focus on the case $n=360$, since ENV and STORE implementations do not run successfully for $n=120$, as noted earlier.

\begin{figure}[!htbp]\centering
    % \hspace{-3cm}
   \includegraphics[width=1.05\textwidth]{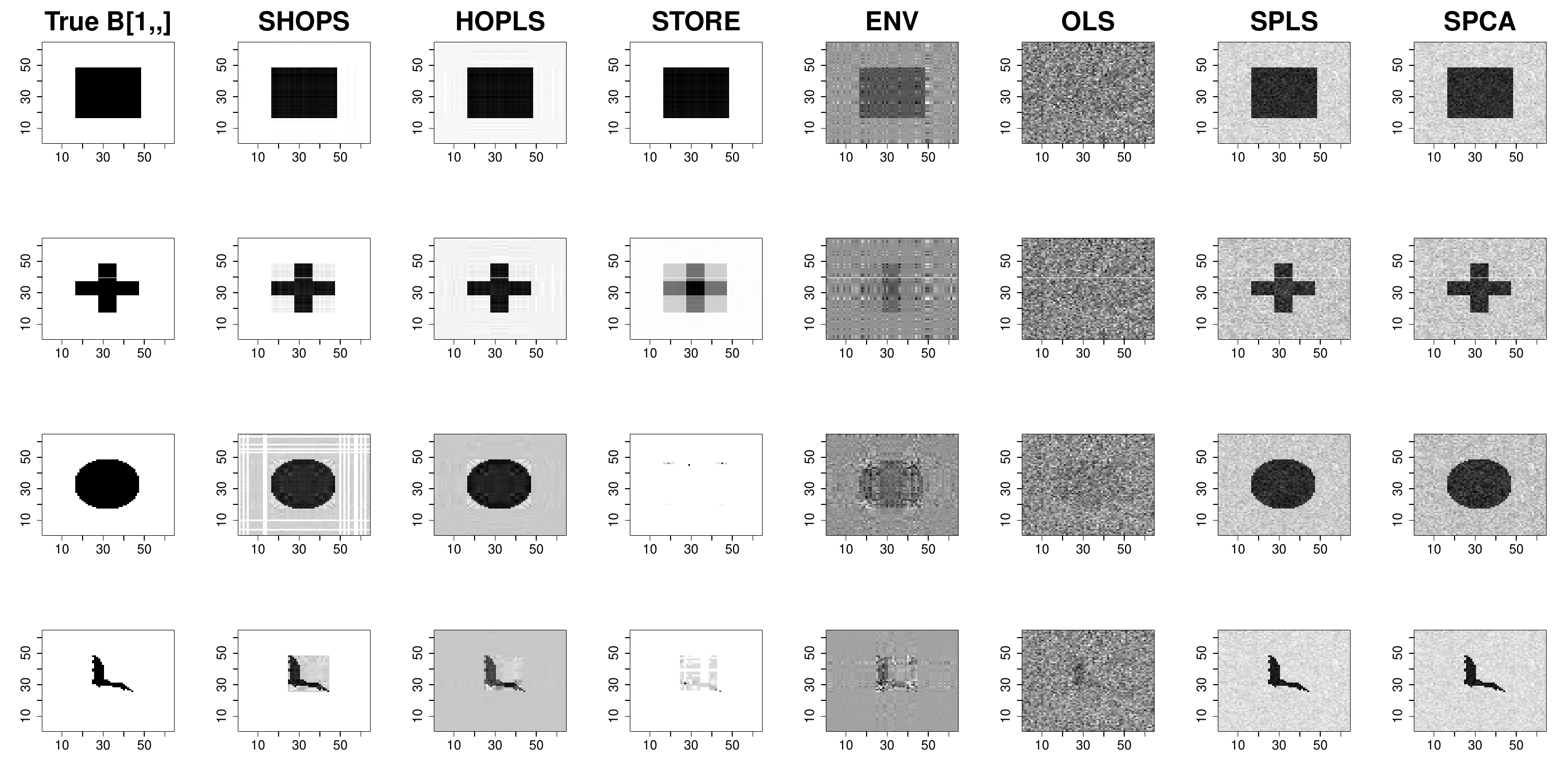}
\caption{ \textbf{Visualization $\tB[1,,]$ and $\hat{\tB}[1,,]$ across methods (columns) and $R\in\{1,2,9,14\}$ (rows).} 
For each $R$, the data was generated from the setting $n=360$, $s=30$. Entries with white color are filled with zeros. The same figure with $s=60$ is displayed in \textbf{Fig.}~\cref{supp-fig:betahats_s=60} }
\label{fig:betahats} 
\end{figure}

We specifically compared these methods based on estimation error and the $\mbox{TPR}_{2,3},\mbox{FPR}_{2,3}$ as displayed in \textbf{Tab.}~\ref{tab:simul_res_R=1_14}. Across all the settings, SHOPS results in the lowest estimation error amongst these methods, which can also be visually seen from \textbf{Fig.}~\ref{fig:betahats}. While HOPLS returns fairly small estimation error, the estimated $\hat{\tB}$ still harbors redundant entries estimated as non-zero, which is the main reason for having higher estimation error than SHOPS. STORE gives lower estimation errors than those of ENV consistently, which is also observed from \cite{sun2017store}, but the values are higher than those of HOPLS. Even if STORE can incorporate sparseness along the second and third modes,  the estimation of the corresponding loadings $\bm{u}_{2r},\bm{u}_{3r}$ relies on the correct estimation of $\bm{u}_{1r}$, so that the estimation performance can be worse than HOPLS and even SPLS. %Note that PLS-type methods can benefit from identifying more relevant latent variables to the response, as discussed in Section~\ref{subsec:varselect_dimred}. 

We next turned our attention to the support recovery performances along the response directions (second and third modes). Among the methods we compared against SHOPS, STORE is the only method that can perform this task. As for the false positive rates represented by $\mbox{FPR}_{2,3}$ in \textbf{Tab.}~\ref{tab:simul_res_R=1_14}, STORE returns exact zero, except for $n=360,R=1,s=60$ case. SHOPS returns $\mbox{FPR}_{2,3}$ fairly close to zero, with the worst value controlled around $0.2$, which is less than that of STORE (0.7). In contrast, the $\mbox{TPR}_{2,3}$ of SHOPS is uniformly better than that of STORE across the settings we considered, while reaching around 0.83 in the worst case ($n=360,R=14$). Collectively, while STORE and SHOPS perform similarly in terms of support recovery along the response directions, STORE tends to be more stringent in identifying active supports compared to SHOPS.

Overall, these simulation results support applicability of SHOPS for the tensor response regression  for achieving   dimension reduction, variable selection as well as  denoising and recovering active entries of the sparse and noisy tensor response $\tY$.

\section{Case study: Application to sn-m3C-seq data of adult mouse hippocampus}

We next applied SHOPS to a
scientific problem that involves
single cell-level chromatin conformation and DNA methylation measurements of the genome as quantified by the  sn-m3C-seq assays \citep{liu2021dna}. 
%This setting includes tensor responses and matrix covariates.
%\subsection{Application to sn-m3C-seq data of adult mouse hippocampus}
sn-m3C-seq  assay is a multi-omics assay that simultaneously profiles   long-range chromatin contacts (i.e., Hi-C for high throughput chromosome conformation capture)  and  DNA methylation (epigenomics)  at the single cell level \citep{liu2021dna}. Both long-range contacts of the genome and DNA methylation are  key components of regulation of gene expression. We showcase application of SHOPS to investigate how long-range chromatin contacts relate to methylation across genomic loci with sn-m3C-seq  data from adult  mouse hippocampus  \citep{liu2021dna}. We refer to the  long-range chromatin contact portion of the data as scHi-C (single cell Hi-C) for short. 

Following the tensor viewpoint of \cite{park2024joint} on scHi-C data, for a single cell $i\in [n]$, we have 20 chromosomes (for mouse genome) and a Hi-C contact matrix of size $d_{2,chr} \times d_{3,chr} $ for each $chr\in [20]$.  Each $(j,k)$-th entry of a contact matrix denotes the contact counts quantifying the physical contact level between genomic loci $j\in [d_{2,chr}]$ and $k\in [d_{3,chr}]$. When the contact matrices are stacked along cells $(n=112)$, the scHi-C data can be viewed as a third order tensor $\tY_{chr}\in \mathbb{R}^{n\times d_{2,chr}\times d_{3,chr}}$ for each chromosome $chr\in [20]$. While the contact matrices traditionally have "bins" (intervals of size 1Kb-1Mb tiling the genome) as rows and columns,  we subset these to include genes and enhancers in order to constrain the second and third modes of the tensor to known functional units of the genome. Specifically,  the second mode of a chromosome-specific tensor  is defined as  the \textit{genes} and the third mode is defined as the \textit{enhancers}, which are typically noncoding regions within $\pm$ 1Mb of the genes and   play essential roles in regulating the genes.   Hence,  $d_{2,chr}$ denotes the number of genes and $d_{3,chr}$ denotes the number of enhancers.  In addition, for a chromosome $chr\in [20]$ within each cell $i \in [n]$, we have DNA methylation level profiled throughout $d_{1,chr}$ genomic loci, i.e., methylation covariate matrix $X_{chr}\in \mathbb{R}^{n\times d_{1,chr}}$. 

The original publication of \cite{liu2021dna} provided cell type labels of the cells profiled. Using this labelling information, we created a  response tensor $\tY_{chr}$ using genes with differential contact levels throughout four cell types: CA3, MGC, OPC, and  VLMC. While our initial set of enhancers harbored the enhancers existing in gene-cCRE (cis-regulatory elements) correlation lists summarized by \cite{li2021atlas},  we further subsetted  these enhancers and kept the ones with differential methylation levels throughout the four cell types. The methylation covariate matrix $X_{chr}$ was also subsetted to include the same set of enhancers. For the analysis, we focused on analyzing  MGC meta cells ($n=112$) generated by aggregating (taking average of) three randomly sampled original MGC cells for both the scHi-C and DNA methylation (Supplementary Section~\cref{supp-preprocess} for further details). 

\noindent \textbf{\textit{SHOPS analysis identifies gene-enhancer regulatory relations.}}
To elucidate  how the DNA methylation level at  enhancers affects the physical contact between genes and enhancers, we fitted the SHOPS model $\tY_{chr}=\tB_{chr}\times_1 X_{chr}+\tE_{chr}$ for each chromosome $chr$ separately. To the best of our knowledge, this work is the first to investigate the impact of methylation on long-range contacts with a tensor regression approach. 
%As far as we know, our paper is the first that applies tensor regression method to this joint analysis of genomic and epigenomic data.
Therefore, we will further elaborate on what the model aims to capture. 

\begin{figure}[!htbp]\centering
    % \hspace{-3cm}
   \includegraphics[width=1.27\textwidth]{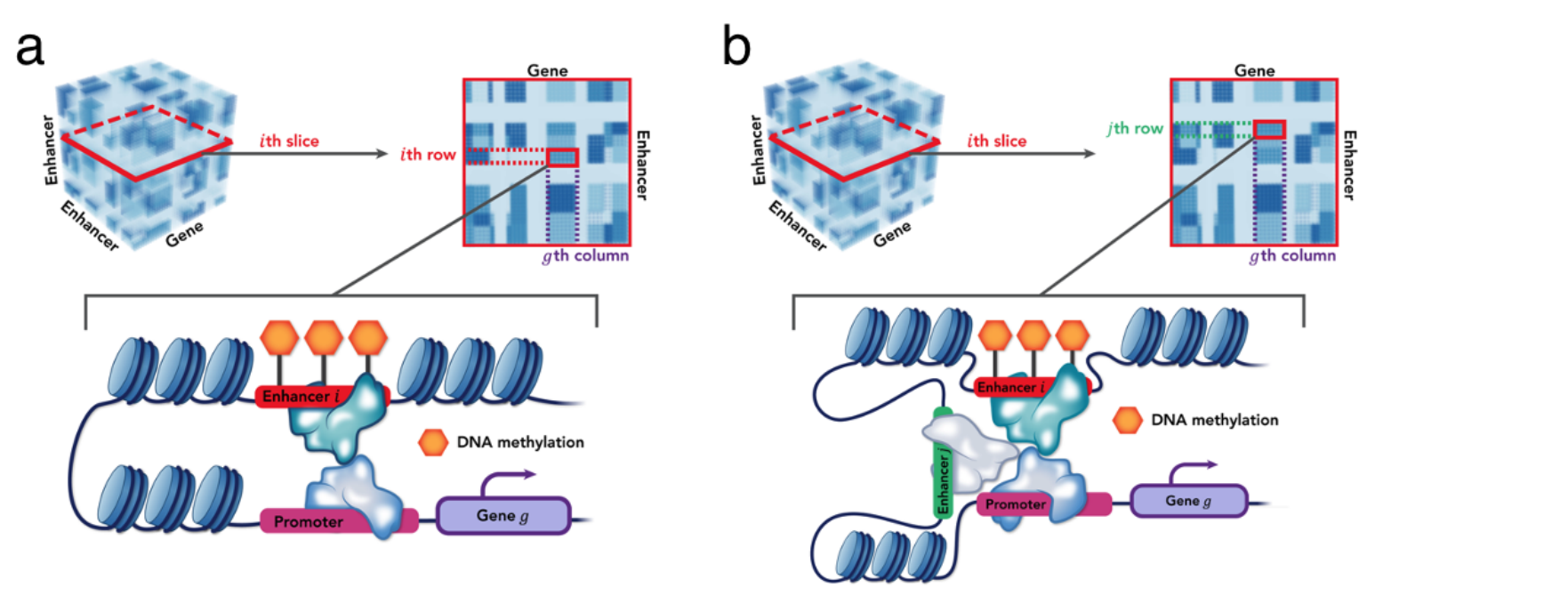}
\caption{Interpretation of entries of the coefficient tensor $\tB_{chr}$  for SHOPS application to  sn-m3C-seq data of \cite{liu2021dna} \textbf{a.} Interpretation of super-diagonal entries of $\tB_{chr}$.  The coefficient tensor $\tB_{chr}$ depicted in the top left panel denotes an \textit{enhancer} ($d_{1,chr}$)$\times$\textit{gene} ($d_{2chr}$)$\times$\textit{enhancer} ($d_{3chr}$) tensor. The $i$th slice of the first mode of the coefficient tensor $\tB_{chr}$ summarizes the effect of DNA methylation at enhancer $i$ on the contact level between the enhancers and the genes on the chromosome $chr$ (top right panel). On the $i$th slice of the coefficient tensor, each entry in the $i$th row depicts the effect of DNA methylation at enhancer $i$ on the contact between the enhancer $i$ and a gene. Specifically, as the bottom panel depicts, DNA methylation (orange hexagons) at  enhancer $i$ (red box) influences the contact between the target gene $g$ (purple box) and enhancer $i$. The square box entry of the top right panel shows the effect size. \textbf{b.} Interpretation of off-diagonal entries of the coefficient tensor $\tB_{chr}$ and multi-way contacts. The top two panels are equivalent to those of \textbf{a}. On the $i$th slice of the coefficient tensor, each entry in the $j$th row depicts the effect of DNA methylation at enhancer $i$ on the contact count between the $j$th enhancer and a gene. Specifically, as the bottom panel shows, DNA methylation (orange hexagons) at an enhancer $i$ (red box) influences the contact between another enhancer $j$ (yellow box) and the target gene $g$ (purple box), which indicates a multi-way contact in the chromatin. The square box entry of the top right panel depicts the effect size.
}
\label{fig:diagonal_between_diagram} 
\end{figure}

\begin{figure}[!htbp]\centering
    % \hspace{-3cm}
   \includegraphics[width=1\textwidth]{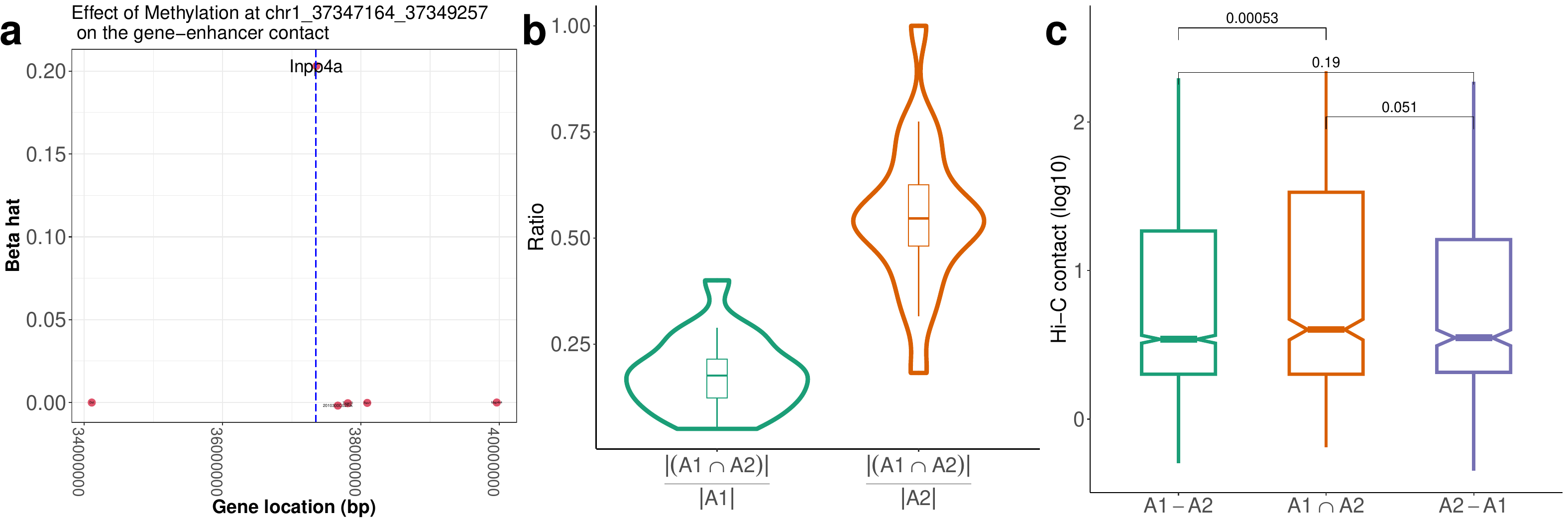}
\caption{Gene regulatory elements and SHOPS estimator. \textbf{a}. Dot plot of $\hat{\tB}_{1[i ,\, ,i]}$ fiber. x-axis denotes the genomic location of genes (10 Mega base pair (Mbp) around the gene), and the y-axis denotes the coefficient value on the fiber, which summarizes the association level between the DNA methylation level at the $i$th enhancer (blue vertical line) and the physical contact of a gene (red dot) and the enhancer $i$. \textbf{b}. Boxplots depicting the overlap rates $|A_1\cap A _2|/|A_1|$ and $|A_1\cap A _2|/|A_2|$. \textbf{c}. Boxplots depicting the scHi-C contact levels for the g-e pairs  in each of $A_1-A_2, A_1\cap A _2,A_2-A_1$. For c,d g-e pairs in all the chromosomes are considered, and for d, the contact levels are normalized so that the contact level is not biased by gene or enhancer length. } 
\label{fig:diagonal} 
\end{figure}

Firstly, we note that for $\tB_{chr}\in\mathbb{R}^{d_{1,chr}\times d_{2,chr}\times d_{3,chr}}$, a tensor fiber $\tB_{chr[i ,\, ,i]}\in\mathbb{R}^{d_{2,chr}}$ represents the effect of DNA methylation at enhancer $i$ on the contact between that enhancer (enhancer $i$) and genes. If  gene $g$ has a large coefficient value $\tB_{chr[i ,g,i]}$, this implies that the methylation of enhancer $i$ affects its physical contact with gene $g$, and that enhancer $i$ is a potential regulator of the gene $g$ (\textbf{Fig.}~\ref{fig:diagonal_between_diagram}a).  \textbf{Fig.}~\ref{fig:diagonal}a visualizes the tensor fiber  $\hat{\tB}_{chr[i ,\, ,i]}$ for $i$-th enhancer located at chr1:37,347,164-37,349,257, and  reveals that methylation at this enhancer affects the physical contact of the enhancer with  gene \textit{Inpp4a} the most, compared to other genes. 
This suggests that the enhancer is a potential regulator of the gene \textit{Inpp4a}, which is proximal to the enhancer in genomic location. SHOPS model facilitates such gene selection by imposing sparsity along the gene mode. 

In order to validate that non-zero entries of the estimated $\tB_{chr[i ,\, ,i]}$ fiber, suggest gene regulation by enhancer $i$, we compared these non-zero entry sets against the external gene-enhancer co-accessibility matrix provided by \cite{li2021atlas}. Gene-enhancer co-accessibility is inferred from another type of single cell profiling assay named sn-ATAC-seq  and is a proxy that is commonly used to link enhancers to genes \citep{pliner2018cicero}. It operates on the assumption that if an enhancer is a regulator of a gene, the enhancer and the promoter region of the gene would exhibit co-accessibility across the cells. However, the reverse is not necessarily true.  
We considered 41,304 g-e (gene-enhancer) pairs  across all chromosomes that were both part of our analysis and in the co-accessibility matrix. Next, we identified the g-e pairs with significant  co-accessibility at false discovery rate (FDR) of 0.05 using the statistical analysis results that were reported by the co-accessibility analysis of \cite{li2021atlas}. We further defined 
 $A_1=\{\mbox{g-e pairs with significant co-accessibility at FDR  0.05}\}$, and $A_2=\{\mbox{g-e pairs with non zero } \hat{\tB}_{chr[i ,\, ,i]}\}$. 
 
 \textbf{Fig.}~\ref{fig:diagonal}b displays the overlap between these two sets across all chromosomes and highlights that majority of the 
 g-e links identified by SHOPS $(A_2)$ have significant g-e co-accessibility at FDR of 0.05, i.e., about $60\%$ of the $A_2$ set g-e pairs are among the $A_1$ set. In contrast, the reverse does not hold, i.e., only a small proportion (15\%) of $A_1$ set g-e pairs are among the $A_2$  set.  This aligns with the intuition that  g-e co-accessibility  does not necessarily imply long-range contacts. Such correlations in co-accessibility could arise due to co-expressed genes (i.e., when two genes are co-expressed, accessibility of their enhancers could correlate with each other's promoter accessibility).  

Next,  to further validate that the g-e pairs within $A_2$ are better supported 
compared to those within $A_1$, we compared their scHi-C contact counts. 
Overall, we would expect that the contact level between a gene and its regulator  is higher than the contacts between the gene and a non-regulating enhancer \citep{miele2008long,schoenfelder2019long}, hence leading to higher contact counts.  \textbf{Fig.}~\ref{fig:diagonal}c yields that the contact counts   of g-e pairs in the intersection,  g-e $\in A_1\cap A_2$, is significantly higher  than those pairs specific to $A_1$, g-e $\in A_1- A_2$ ($\mbox{p-value}<5.3e-04$), which consolidates the results of the co-accessibility analysis. 
%This shows that the $g-e$ pairs, whose contact level is highly affected by the methylation level happening at the enhancer and whose co-accessibility level is high, are prone to have significantly higher physical contact level than the others. 
%Furthermore, we also observe that contact counts  for g-e pairs that are specific to $A_2$  (g-e $\in  A_2-A_1$) are also higher than those pairs specific to $A_1$, further consolidating the results of the co-accessibility analysis.    
%set is significantly lower than the other groups. Hence, the results represent that the co-accessibility level alone does not contain sufficient enough information on gene regulations by enhancer, while the SHOPS result better aligns with the gene regulatory pattern. The alignment is even better when we fuse the SHOPS result with the co-accessibility results as the highest contact level at $A_1\cap A_2$ shows \textbf{Fig.}~\ref{fig:diagonal}d.

%\label{fig:between_diagram} 
%\label{fig:diagonal_diagram} 

\begin{figure}[!htbp]\centering
    % \hspace{-3cm}
   \includegraphics[width=1\textwidth]{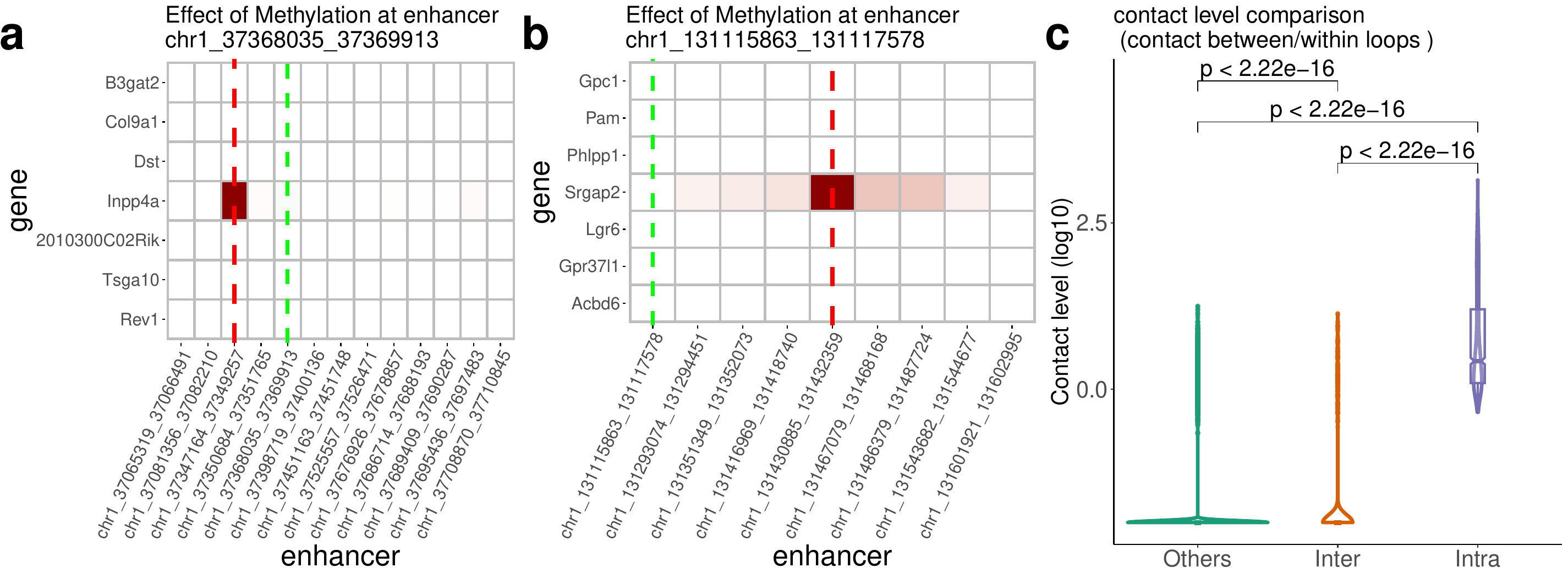}
\caption{\textbf{a.b,} SHOPS estimates for the coefficient slice matrix $\tB_{1[i ,\, ,\,]}$. x-axis denotes  the enhancers within 1Mbp region around the $i$th enhancer (a. chr1:37,368,035-37,369,913, b. chr1:131,115,863-131,117,578) and y-axis lists the nearby genes. Each entry of the matrix denotes the estimated coefficient value with greener (redder) values correspond to larger positive (negative) values. Green vertical line denotes the location of the enhancer $i$, red vertical line denotes the location of enhancer with  the maximum estimated coefficient magnitude in the matrix. \textbf{c}. scHi-C contact levels (normalized as in \textbf{Fig.}~\ref{fig:diagonal}) of the three groups of genomic regions \textbf{Intra:} gene-enhancer pairs with non-zero estimated $\tB_{chr[i ,\, ,i]}$ values for all $i$, $chr$. \textbf{Inter:} enhancer$_i$-enhancer$_j$ pairs, where enhancer$_j$ has the maximum  effect size in absolute value in the estimated  $\tB_{chr[i ,\, ,\,]}$ matrix for each $i$, $chr$. \textbf{Others:} enhancer-enhancer pairs excluding Inter group pairs. Since the Intra group has disproportionately large contact levels, all the contact levels are transformed with $-\log_{10}(x+0.01)$ to visualize the three groups in one plot. Each pairs of groups is compared with the  Wilcoxon rank sum test.
 } \label{fig:between} 
\end{figure}

\noindent \textbf{\textit{SHOPS analysis identifies multi-way regulatory relations.}}
Thus far, we evaluated the SHOPS results in terms of the impact of methylation of an enhancer to that enhancer's contacts with the genes. However,  SHOPS analysis also generates candidates of multi-way contacts (i.e., also known as many-body interactions \citep{perez2020chromatix,liu2021extracting}) in the genome. Such multi-way contacts arise when a gene is regulated by multiple interacting enhancers. The  SHOPS coefficient entry $\tB_{chr[i ,g,j]}$ can be interpreted as the effect of methylation  at $i$th enhancer on the contact count between gene $g$ and enhancer $j$ (\textbf{Fig.}~\ref{fig:diagonal_between_diagram}b). For example, \textbf{Fig.}~\ref{fig:between}a,b display an estimated tensor slice $\hat{\tB}_{chr[i ,\, ,\,]}$ for two enhancers (a for enhancer located at chr1:37,368,035-37,369,913, b for enhancer located at chr1:131,115,863-131,117,578). \textbf{Fig.}~\ref{fig:between}a yields that the methylation  at  enhancer located at chr1:37,368,035-37,369,913 (green vertical line) affects the contact between gene \textit{Inpp4a} and the enhancer 
chr1:37,347,164-37,349,257 (red rectangular box). %This could represent a possibility of a clique structure between the methylated enhancer, the target gene, and the other enhancer for the gene.
While these two enhancers in \textbf{Fig.}~\ref{fig:between}a are close to each other in their genomic coordinates, 
the three-way contact could also occur among enhancers  that are distal to each other (\textbf{Fig.}~\ref{fig:between}b).

As we demonstrated previously, the g-e pairs with non-zero estimated coefficients in $\hat{\tB}_{chr[i ,\, ,i]}$ have significantly higher contact counts compared to the others (boxplot for \textbf{Intra} group in \textbf{Fig.}~\ref{fig:between}c).  To further validate enhancer-enhancer contacts in the inferred  g-e-e triplet (gene $g$, enhancer $i$, enhancer $j$), we quantified the  contact counts between the two enhancers ($i$ and $j$) (boxplot across chromosomes  for \textbf{Inter} group in \textbf{Fig.}~\ref{fig:between}c) and observed that these enhancers harbor significantly more contact counts than  enhancers that are not inferred to interact by SHOPS
(Wilcoxon ranksum test p-value of $2.2e-16$ with Benjamini-Hochberg \citep{benjamini1995controlling} correction).
%Furthermore, we observe   that the interactions between two enhancers, one of whose methylation affects the other's physical contact with its target gene (Boxplot for \textbf{Inter} group in \textbf{Fig.}~\ref{fig:between}b), is also significantly (Wilcoxon ranksum test $\mbox{p-value}=0.035$ with Benjamini-Hochberg correction) higher than the contact between two random enhancers without such interaction derived from SHOPS estimator (Boxplot for \textbf{Other} group in \textbf{Fig.}~\ref{fig:between}b). 
In summary, SHOPS applied to sn-m3C-seq data has the potential to reveal both enhancers of genes and multi-way contacts involving multiple enhancers.

\section{Discussion}\label{sec:discusion}

In this paper, we presented SHOPS model for simultaneous dimension reduction of the high-dimensional predictor variables, selection of important variables, and tensor response denoising. Theoretical investigations of the SHOPS model and the simulation studies illustrated  that SHOPS yields consistent estimator $\hat{\tB}_{SHOPS}$ under the high-dimensional regime. Application of SHOPS to a real dataset highlighted the utility of SHOPS for modern high-dimensional tensor response regression problems. Specifically, the application to sn-m3C-seq \citep{liu2021dna} data revealed SHOPS as identifying regulatory elements of genes and multi-way interactions among genes and two different enhancers. The latter is a relatively less  explored area of single cell genomics.

This work establishes sparse tensor response partial least squares method as a modeling approach in the high-dimensional tensor response and matrix covariate settings. Moreover, it also addresses the consistency rates of the  tensor (multivariate) response (both sparse or non-sparse) partial least squares regression when both the dimensions on response tensor and the numbers of predictors increase. While \cite{chun2010sparse} and \cite{zhu2020envelope} provide consistency result of the multivariate response partial least squares, they rely on the assumption that the dimensionality of the response matrix is fixed, which is a restrictive assumption in modern applications arising from single cell genomics. 
%In addition to the SHOPS,  we also devised two baseline algorithms: SHOSVD-PLS (Algorithm~\cref{supp-alg:SHOSVD-PLS}) and SHOOI-PLS (Algorithm~\cref{supp-alg:SHOOI-PLS}). These algorithms are direct tensor extensions of the SPLS algorithm of \cite{chun2010sparse}. Interestingly,  SHOOI-PLS empirically achieves comparable performance to SHOPS, even though it lacks theoretical justification.

While this paper provides theoretical guarantees of the sparse higher order partial least squares model, existing proposition about the equivalence between PLS and envelope regression by \cite{cook2013envelopes} implies that an envelope extension of the SHOPS model could potentially enable statistical inferences. In addition, we employed Gaussian noise model on the response $\tY$, but a natural extension to the sub-Gaussian GLM model can be made similar to the extension obtained by \cite{chung2010sparse}. Lastly, inference on other parameters of the response tensor could also be obtained. For example, we expect that the PLS or envelope framework on the quantile regression \citep{ding2021envelope} and tensor quantile regression \citep{li2021tensor} joined together could make SHOPS applicable for quantile estimation of the high-dimensional response variables.

\if1\blind
{
\section{Funding}
This work is supported in part by funds from the National Institutes of Health (NIH: R01HG003747, R21HG012881) and Chan Zuckerberg  Initiative Data Insights award.}\fi

\section{Disclosure statement}
The authors report there are no competing interests to
declare.

\bibliographystyle{chicago}

\bibliography{bibtex}
\end{document}